
\input psfig
\input mn.tex

\def\vc{v_{\rm c}}
\def\onetwo{{\textstyle {1 \over 2} \displaystyle}}

\def\atan{{\rm atan}}

\def\acos{{\rm acos}}
\def\cosec{{\rm cosec}}

\def\sech{{\rm sech}}
\def\csch{{\rm csch}}

\def\data{{\rm data}}
\def\model{{\rm model}}
\def\rmax{r_{\rm max}}

\def\kms{kms$^{-1}$}
\def\as{a_{\rm s}}
\def\ra{r_{\rm a}}
\def\vlosi{v_{\odot,i}}
\def\vlos{v_{\odot}}
\def\vlosic{v_{r\odot,i}}
\def\vlosc{v_{r\odot}}

\def\vT{v_{\rm t}}
\def\rhos{\rho_{\rm s}}

\def\sM{s_{\rm M}}

\def\Msun{{\rm M}_\odot}
%
%
\def\spose#1{\hbox to 0pt{#1\hss}}
\def\lta{\mathrel{\spose{\lower 3pt\hbox{$\sim$}}
    \raise 2.0pt\hbox{$<$}}}
\def\gta{\mathrel{\spose{\lower 3pt\hbox{$\sim$}}
    \raise 2.0pt\hbox{$>$}}}
\def\today{\ifcase\month\or
 January\or February\or March\or April\or May\or June\or
 July\or August\or September\or October\or November\or December\fi
 \space\number\day, \number\year}
\newdimen\hssize
\hssize=8.4truecm  
\newdimen\hdsize
\hdsize=17.7truecm    


\newcount\eqnumber
\eqnumber=1
\def\chaphead{}
 
\def\new{\hbox{(\rm\chaphead\the\eqnumber)}\global\advance\eqnumber by 1}
 
\def\first{\hbox{(\rm\chaphead\the\eqnumber a)}\global\advance\eqnumber by 1}
\def\last#1{\advance\eqnumber by -1 \hbox{(\rm\chaphead\the\eqnumber#1)}\advance
     \eqnumber by 1}
 
\def\ref#1{\advance\eqnumber by -#1 \chaphead\the\eqnumber
     \advance\eqnumber by #1}
 
\def\nref#1{\advance\eqnumber by -#1 \chaphead\the\eqnumber
     \advance\eqnumber by #1}

\def\eqnam#1{\xdef#1{\chaphead\the\eqnumber}}
 


\newcount\tabnumber 
\tabnumber=1
\def\tabnew{\global\advance\tabnumber by 1}
\def\tabnam#1{\xdef#1{\chaphead\the\tabnumber}}


\newcount\fignumber 
\fignumber=1
\def\fignew{\global\advance\fignumber by 1}
\def\fignam#1{\xdef#1{\chaphead\the\fignumber}}


\pageoffset{-0.85truecm}{-1.05truecm}



\pagerange{}
\pubyear{version: \today}
\volume{}


\begintopmatter

\title{The Mass of the Andromeda Galaxy}

\author{N.W.\ Evans$^1$ and M.I.\ Wilkinson$^{1,2}$}

\vskip0.15truecm
\affiliation{$^1$Theoretical Physics, Department of Physics, 1 Keble Road,
                 Oxford, OX1 3NP}
\vskip0.15truecm

\affiliation{$^2$Institute of Astronomy, Madingley Rd, Cambridge, CB3
0HA}

\shortauthor{N.W.\ Evans and M.I.\ Wilkinson} 

\shorttitle{The Mass of the Andromeda Galaxy}



\abstract{This paper argues that the Milky Way galaxy is probably
the largest member of the Local Group.  The evidence comes from
estimates of the total mass of the Andromeda galaxy (M31) derived from
the three dimensional positions and radial velocities of its satellite
galaxies, as well as the projected positions and radial velocities of
its distant globular clusters and planetary nebulae. The available
dataset comprises ten satellite galaxies, seventeen distant globular
clusters and nine halo planetary nebulae with radial velocities. We
find the halo of Andromeda has a mass of $\sim 12.3^{+18}_{-6} \times
10^{11}\Msun$, together with a scalelength of $\sim 90$ kpc and a
predominantly isotropic velocity distribution.  For comparison, our
earlier estimate for the Milky Way halo is $19^{+36}_{-17} \times
10^{11}\Msun$. Though the error bars are admittedly large, this
suggests that {\it the total mass of M31 is probably less than that of
the Milky Way}. We verify the robustness of our results to changes in
the modelling assumptions and to errors caused by the small size and
incompleteness of the dataset.

Our surprising claim can be checked in several ways in the near
future.  The numbers of satellites galaxies, planetary nebulae and
globular clusters with radial velocities can be increased by
ground-based spectroscopy, while the proper motions of the companion
galaxies and the unresolved cores of the globular clusters can be
measured using the astrometric satellites {\it Space Interferometry
Mission} (SIM) and {\it Global Astrometric Interferometer for
Astrophysics} (GAIA).  Using 100 globular clusters at projected radii
$20\lta R\lta 50$ kpc with both radial velocities and proper motions,
it will be possible to estimate the mass within $50$ kpc to an
accuracy of $\sim 20\%$.  Measuring the proper motions of the
companion galaxies with SIM and GAIA will reduce the uncertainty in
the total mass caused by the small dataset to $\sim 22\%$.}


\maketitle  

%
\section{Introduction}
The aim of this paper is to obtain an accurate estimate of the total
mass of the halo of the Andromeda galaxy (M31). The best probes of the
mass distribution at large distances are the satellite galaxies and
the distant globular clusters. The gas rotation curve can only be
tracked out to $\sim 30$ kpc (e.g., Roberts \& Whitehurst 1975, Hodge
1992), and so -- as is the case for the Milky Way -- the most
substantial clues as to the large-scale mass distribution come from
statistical analyses of the satellite kinematics.

The Local Group has two prominent subgroups of satellites centered on
its two largest members, the Milky Way and the Andromeda galaxies
(e.g. van den Bergh 1999a).  A number of authors have
already studied the problem of estimating the mass of the Milky Way
from its satellite subgroup (e.g., Little \& Tremaine 1987; Kulessa \&
Lynden-Bell 1992; Kochanek 1996). The most recent estimate by
Wilkinson \& Evans (1999, hereafter WE99) found a total mass of $\sim
1.9 \times 10^{12} \Msun$ and an extent of $\sim 170$ kpc. By
contrast, until recently, the companion problem of estimating the mass
of the Andromeda galaxy from its satellite subgroup has received
hardly any attention. Hodge (1992) lists a number of determinations of
the mass of M31, but almost all use either the optical and radio
rotation curves (e.g., Rubin \& Ford 1970) or the inner globular
clusters (e.g., Hartwick \& Sargent 1974; van den Bergh 1981) and so
are really measurements of the mass within $\sim 30$ kpc. Recently
Courteau \& van den Bergh (1999) estimated the mass of the M31 halo
using only seven satellite galaxies. An analysis of all available data
on objects outside $\sim 20$ kpc has not been carried out to date.

It is especially timely to look at the problem of determining the mass
and the extent of the Andromeda halo now. First, the last few years
have seen the discovery of a number of faint dwarf spheroidal
companions of Andromeda (see e.g. Armandroff \& Da Costa 1999) as a
result of two ongoing surveys of the sky around Andromeda. To date,
one of these surveys (Armandroff \& Da Costa 1999) has covered 1550
square degrees while the second (Karachentsev \& Karachentseva 1999)
has scanned a circular area of radius 22 degrees centred on M31. The
quantitative completeness limits of these surveys are not yet
available. Armandroff \& Da Costa (1999) claim that their detection of
And V with absolute $M_V \sim -10.2$ is highly significant and
maintain that they have the sensitivity to detect still fainter
objects, such as dwarf spheroidals with $M_V \sim -8.5$.
Complementing these programmes are the ongoing radial velocity surveys
of M31 globular clusters (Perrett et al. 1999) and halo planetary
nebulae (Ford et al. 1989). The dataset is likely to increase
substantially in richness over the next few years, suggesting that
more detailed models of the dynamics of the outer parts of M31 are
warranted.  Second, a number of groups are conducting pixel lensing
experiments towards Andromeda (e.g., Crotts \& Tomaney 1996; Kerins et
al. 2000). The lenses may lie in the halos of the Milky Way and M31,
as well as the disk of the Milky Way and the bulge and disk of
M31. Fortunately, there is a possible diagnostic of microlensing by
M31's halo. As the galaxy is highly inclined, lines of sight to the
far side pass through more of the M31 halo than those to the near side
(Crotts 1992; Kerins et al. 2000). However, the amplitude of this
near-far disk asymmetry depends on how extensive and massive the halo
of Andromeda really is (Evans \& Wilkinson 2000).

In Section 2, we outline the properties of our halo model, deriving
both projected properties and simple distribution functions
(DFs). Section 3 describes our models for the satellite galaxies, the
globular clusters and the planetary nebulae, as well as the available
dataset. In Section 4, the mass estimator algorithm of Little \&
Tremaine (1987) is adapted to the case when only projected data are
available. For the satellite galaxies, the three dimensional position
with respect to Andromeda's centre and the line of sight velocities
are known. For the globular clusters and halo planetary nebulae, only
the projected positions and line of sight velocities are
available. The next two sections present our analysis of the data and
estimates of the mass and extent of the Andromeda galaxy halo,
together with the errors caused by incompleteness and small number
statistics. In Section 7, we summarise our conclusions and describe
the prospects for the future.
\section{A Model of the Andromeda Halo}
\subsection{The Density and Projected Mass}
A simple representation of the Andromeda halo has a roughly flat
rotation curve out to an unknown cut-off. The TF (or truncated, flat
rotation curve) model examined in detail by WE99 has exactly these
properties, as well as the virtue of analytical simplicity. The
potential-density pair is
\eqnam{\potdens}
$$\eqalign{\rho(r) =& {M\over 4\pi}{a^2\over r^2 (r^2 + a^2)^{3/2}},\cr
\psi(r) =& v_0^2\log\Bigl({\sqrt{r^2 + a^2} + a \over r}\Bigr),\cr}
\eqno\new$$
where $M$ is the total mass of the halo and $a$ is a measure of the
extent. The rotation curve is flat with amplitude $v_0 = \sqrt{GM/a}$
in the inner parts. In modelling the halo of Andromeda, we assume $v_0
= 240$ \kms.  This gives a circular velocity of $235$ \kms\ at a
radius of $30$ kpc which is in agreement with Hodge (1992, chapter 7).

As we observe M31 in projection, it is helpful to have available the
projected properties of our halo model.  Mateo (1998), following
Karachentsev \& Makarov (1996), estimates the distance $D$ of M31 as
770 kpc, although there is still some uncertainty as to this value
(c.f. Feast 1999).  For an observer at finite distance $D$, the
projected surface density $\Sigma$ is
$$\Sigma(R) = \biggl[2\int_R^D dr + \int_D^\infty dr \biggr]{r 
\over \sqrt{r^2 - R^2}} \rho(r),\eqno\new$$
from which we obtain
$$\eqalign{\Sigma(R) =& {M \over 4\pi a} \Bigl[\Bigr. {1\over R}
\Bigl( \atan \Bigl({a \over R} {\sqrt{1 - R^2/ D^2}
\over \sqrt{1 + a^2/D^2}}\Bigr) + \atan
\Bigl( {a\over R} \Bigr) \Bigr)\cr  &- {a \over
(R^2 + a^2) }\Bigl( 1 + {\sqrt{1 - R^2/D^2} 
\over \sqrt{1 + a^2/D^2}}\Bigr)\Bigr].} \eqno\new$$
The usefulness of our model (\potdens) derives from the fact that the
density can be written in terms of the potential as
\eqnam{\rhophi}
$$\rho(\psi) = {M\over 4\pi a^3} {\sinh^5 (\psi/v_0^2) \over
\cosh^3 (\psi /v_0^2)}.\eqno\new$$
This is the crucial formula for obtaining distribution functions (DFs)
which are analytically tractable. This is the problem to which we now
turn.
\subsection{The Velocity Distributions}

As we wish to be certain of the robustness of our results, we will
analyse the dataset with two different kinds of velocity
distributions.  A first possibility is to use the ansatz (e.g.,
H\'enon 1973, WE99)
\eqnam{\dejonghe}
$$F(\varepsilon,l) = l^{-2 \beta} f(\varepsilon),\eqno\new$$
where
\eqnam{\henonDF}
$$\eqalign{f(\varepsilon) = & {2^{\beta - 3/2} \over \pi^{3/2} 
\Gamma [m - 1/2 + \beta] \Gamma [1 - \beta]} \cr  
& \times {d \over d\varepsilon} \Bigl[ \int_o^\varepsilon 
d \psi {d^m r^{2\beta} \rho \over d \psi^m} (\varepsilon - \psi)^{\beta - 3/2 + m}
\Bigr].}
\eqno\new$$ 
Here, $m$ is an integer whose value is chosen such that the integral
in (\henonDF) converges. For such a DF, the velocity dispersions
$\langle v^2_{\phi}\rangle$ and $\langle v^2_{\theta}\rangle$ are
equal, and there is a constant orbital anisotropy $\beta = 1 -
\langle v_{\theta}^2\rangle/\langle v_r^2 \rangle$. The properties
of these DFs are discussed fully in WE99.

A second possibility is to assume that the DFs depend on the binding
energy $\varepsilon$ and the angular momentum per unit mass $l$
through the ansatz (Osipkov 1979; Merritt 1985)
\eqnam{\osipmer}
$$F(\varepsilon,l) = f(Q),\qquad\qquad Q = \varepsilon - {l^2\over2\ra^2},
\eqno\new$$
where $\ra$ is the anisotropy length-scale. These DFs have the
property that for $r < \ra$, the velocity distribution is isotropic
while for $r > \ra$ it tends towards radial anisotropy. For any
population with density $\rho$, the Osipkov-Merritt DF is given by
\eqnam{\anisoDF}
$$f(Q) = {1 \over 2\sqrt{2} \pi^2}\int_0^Q {d\psi \over
\sqrt{Q-\psi}}{d^2\over d\psi^2}\left[\left(1+{r^2 \over
\ra^2}\right)\rho(r)\right].\eqno\new$$
Inserting (\rhophi) into (\anisoDF), we obtain
\eqnam{\TFosipmer}
$$\eqalign{f(Q) &= {M \over 2 \sqrt{2}\pi^3 a^3 v_0^3}\int_0^{Q/v_0^2}
{d\phi \over \sqrt{Q/v_0^2-\phi}} \cr
&\times \Biggl[\tanh\phi\sinh^2\phi +\left(1-{3a^2 \over
2\ra^2}\right) \tanh^3\phi \cr 
&+ 3\left(1-{a^2 \over \ra^2}\right)\tanh^3\phi \,\sech^2\phi +
{3a^2 \over 2\ra^2}\tanh\phi \Biggr].}
\eqno\new$$
The integral can be easily evaluated using Gaussian quadrature and
Fig.~1 presents examples of these DFs for several values of the
anisotropy radius $\ra$. The DF is everywhere positive provided that
$\ra \geq 0.092 a$. 

The radial velocity dispersion $\langle v^2_{r}\rangle$ is given by
\eqnam{\anisodisp}
$$\eqalign{\langle v^2_{r}\rangle =  &{v_0^2 (r^2 + a^2)^{1/2} \over 2 a^5 (1 +
r^2/\ra^2)}\big( 2 a^2 r^2 + a^4\left(1 - {r^2\over
\ra^2}\right)\cr & + r^2(a^2+r^2)\left({a^2 \over \ra^2} -
2\right)\log\left[1+{a^2 \over r^2}\right]\big).} \eqno\new$$
As $r \rightarrow \infty$, $\langle v^2_{r}\rangle \rightarrow 0$ and as $r
\rightarrow 0$, $\langle v^2_{r}\rangle \rightarrow v_0^2/2$. The 
anisotropy parameter $\beta = 1 - \langle v^2_{\theta}\rangle/\langle
v^2_{r}\rangle$, which is a measure of the anisotropy of the velocity
ellipsoid at radius $r$, is related to $\ra$ via
\eqnam{\anisoeom}
$$ \beta(r) = {r^2 \over \ra^2+r^2}.\eqno\new$$
This result (\anisoeom) holds true for any DFs of the Osipkov-Merritt
form.
\beginfigure{\fignumber}
\fignam{\figOMDF}
\centerline{\psfig{figure=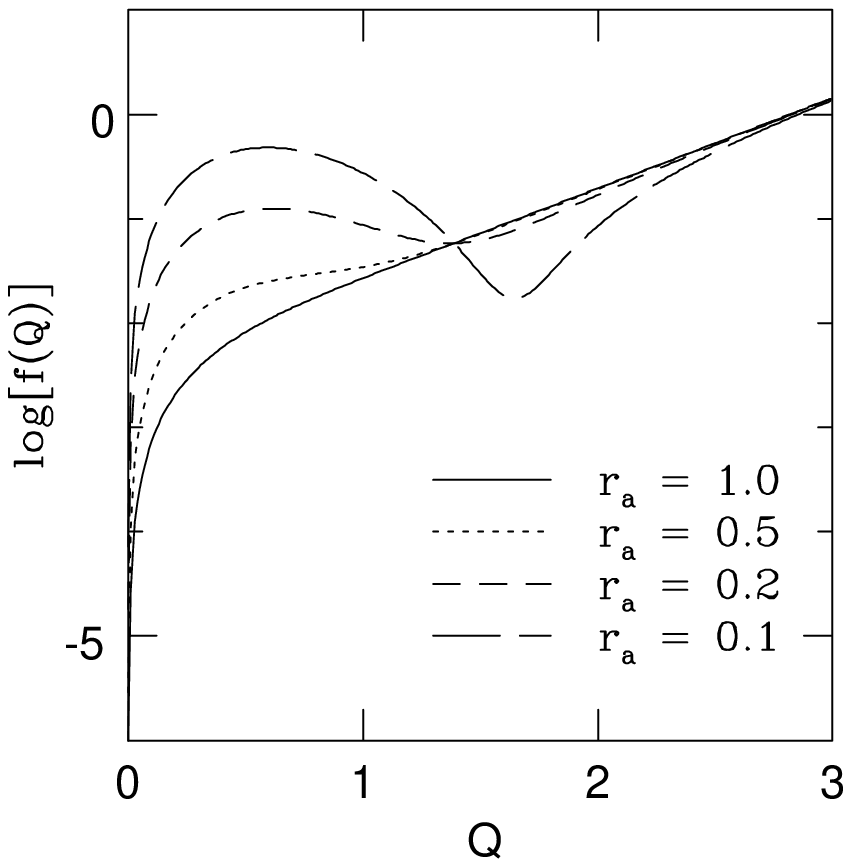,width=\hssize}}
\smallskip\noindent
\caption{{\bf Figure \figOMDF.} Osipkov-Merritt type DFs for a TF model
potential. A range of values of $\ra$ are plotted. Units with
$G=M=a=1$.}
\fignew
\endfigure
\begintable*{\tabnumber}
\tabnam{\satdata}
\def\tableunderline{\noalign{\vskip0.1truecm}\noalign{\vskip0.1truecm}}
\caption{{\bf Table \satdata.} Data on the radial velocities of M31 and its
dwarf satellites.  Listed are Galactic coordinates ($\ell,b$) and
heliocentric distances ($s$ and $\sM$) in kpc, as given by Courteau \&
van den Bergh (1999) and by Mateo (1999), respectively.  Also given
are the observed heliocentric radial velocities ($\vlos$) in
kms$^{-1}$ together with actual and projected distances from the
centre of M31 ($r$ and $R$) in kpc (using the distances of Courteau \&
van den Bergh 1999), corrected heliocentric radial velocities
($\vlosc$) in \kms (in which the solar motion within the Milky Way
and the radial motion of the Milky Way towards M31 is removed) and
object type. See Section~5.1 for a discussion of the velocity
corrections. For M31, the distance is taken from Karachentsev \&
Makarov (1996), while the radial velocity is from Courteau \& van den
Bergh (1999).  For the satellites, all data in columns 3, 4, 5 and 7
are from Courteau \& van den Bergh (1999) with the exception of the
radial velocity of And II which is given by C$\hat{\rm o}$t$\acute{\rm
e}$ et al. (1999).  }
\halign{#\hfil&\quad\hfil#\hfil&\quad\hfil#\hfil&\quad\hfil#\hfil&\quad\hfil#\hfil&\quad\hfil#\hfil&\quad\hfil#\hfil&\quad\hfil#\hfil&\quad\hfil#\hfil&\quad\hfil#\hfil&\quad#\hfil\cr
\noalign{\hrule}
\noalign{\vskip0.1truecm}
Name& Alias & $\ell$& $b$& $s$& $\sM$ & $\vlos$& $r$ & $R$ &
$\vlosc$ & Type\cr
\tableunderline
M31 & NGC 224 & 121.2 & -21.6 & 770 & - & -301 & - & -  & - & SbI-II\cr
\tableunderline
M32 & NGC 221 & 121.2 & -22.0 & 760 & 805 $\pm$ 35 & -205 $\pm$ 3 & 11
& 5 & 95 & E2\cr \tableunderline
NGC 205 & M110 & 120.7 & -21.1 & 760 & 815 $\pm$ 35 & -244 $\pm$ 3 & 13
& 8 & 58 & dSph\cr \tableunderline
NGC 147 & UGC 326 & 119.8 & -14.3 & 660 & 725 $\pm$ 45 & -193 $\pm$ 3 &
144 & 100 & 118 & dSph/dE5\cr \tableunderline
NGC 185 & UGC 396 & 120.8 & -14.5 & 660 & 620 $\pm$ 25 & -202 $\pm$ 7 &
141 & 95 & 107 & dSph/dE3\cr \tableunderline
M33 & NGC 598 & 133.6 & -31.5 & 790 & 840 & -181 &
203 & 198 & 72 & ScII-III\cr \tableunderline
IC 10 & UGC 192 &119.0 & -3.3 & 660 & 825 $\pm$ 50 & -344 $\pm$ 5 & 253
& 243 & -29 & dIrr\cr \tableunderline
And II & - & 128.9 & -29.2 & 700 & 660 $\pm$ 100 & -188 $\pm$ 3 & 149 & 138
& 82 & dSph\cr \tableunderline
LGS3 & Pisces & 126.8 & -40.9 & 810 & 810 $\pm$ 60 & -286  $\pm$ 4 &
276 & 262 & -38 & dIrr/dSph\cr \tableunderline
Pegasus & DDO 210 & 94.8 & -43.5 & 760 & 955 $\pm$ 50 & -182 $\pm$ 2 &
409 & 399 & 86 & dIrr/dSph\cr \tableunderline
IC 1613 & DDO 8& 129.7 & -60.6 & 720 & 700 $\pm$ 35 & -232 $\pm$ 5 &
505 & 489 & -58 & IrrV\cr \tableunderline
\noalign{\vskip0.1truecm}\noalign{\hrule}
}
\tabnew
\endtable
\begintable{\tabnumber}
\tabnam{\gcdata}
\def\tableunderline{\noalign{\vskip0.1truecm}\noalign{\vskip0.1truecm}}
\caption{{\bf Table \gcdata.} Data on the radial velocities of the
outer globular clusters of M31. Listed are Galactic coordinates
($\ell,b$), observed heliocentric radial velocities ($\vlos$) in
kms$^{-1}$ together with projected distances $R$ from the centre of
M31 in kpc and corrected heliocentric radial velocities ($\vlosc$) in
\kms (in which the solar motion within the Milky Way and the radial
motion of the Milky Way towards M31 is removed). See Section~5.1 for a
discussion of the velocity corrections. Sources: (a) Federici et
al. (1993) (b) Kent, Huchra \& Stauffer (1989) (c) Perrett et
al. (private communication) (d) Barmby et al. (2000). Note: The
designation S refers to the catalogue of Sargent et al. (1977)}
\halign{#\hfil&\quad\hfil#\hfil&\quad\hfil#\hfil&\quad\hfil#\hfil&\quad\hfil#\hfil&\quad\hfil#\hfil\cr
\noalign{\hrule}
\noalign{\vskip0.1truecm}
Name&$\ell$&    $b$& $\vlos$ & $R$ & $\vlosc$\cr
\tableunderline
S219$^d$ & 121.2 & -23.0 & -315  $\pm$ 2& 19.5 & -17\cr \tableunderline
S352$^b$ & 122.7 & -21.2 & -325  $\pm$ 20& 19.5 & -27\cr \tableunderline
S9$^d$ & 119.9 & -20.7 & -215  $\pm$ 26& 19.6 & 89\cr \tableunderline
S14$^c$ & 119.9 & -22.5 & -423  $\pm$ 12& 19.7 & -121\cr \tableunderline
S22$^c$ & 120.2 & -22.7 & -374  $\pm$ 12& 19.7 & -73\cr \tableunderline
S343$^c$ & 122.3 & -20.4 & -253  $\pm$ 12& 20.4 & 47\cr \tableunderline
S13$^b$ & 120.0 & -20.5 & 47  $\pm$ 38& 20.4 & 351\cr \tableunderline
S268$^d$ & 121.5 & -20.1 & -321  $\pm$ 26& 20.6 & -19\cr \tableunderline
BA311$^c$ & 122.4 & -20.5 & -97  $\pm$ 12& 21.3 & 202\cr \tableunderline
S327$^c$ & 122.0 & -20.1 & -258.5  $\pm$ 12& 22.3 & 42.2\cr \tableunderline
S3$^a$ & 119.4 & -20.7 & -87  $\pm$ 30 & 24.8 & 218\cr \tableunderline
S353$^a$ & 122.7 & -20.3& -296  $\pm$ 30 & 26.0 & 3\cr \tableunderline
EX8$^d$ & 123.3 & -21.3 & -154 $\pm$ 40 & 26.7 & 142\cr \tableunderline
S339$^a$ & 122.2 & -19.7 & 33  $\pm$ 30 & 28.3 & 333\cr \tableunderline
S355$^a$ & 123.0 & -22.9 & -60 $\pm$ 30 & 28.6 & 235\cr \tableunderline
S2$^d$ & 119.2 & -23.2 & -340  $\pm$ 22& 33.2 & -38\cr \tableunderline
S1$^d$ & 119.0 & -23.2 & -332  $\pm$ 3& 34.1 & -29\cr \tableunderline
\noalign{\vskip0.1truecm}\noalign{\hrule}
}
\tabnew
\endtable
\begintable{\tabnumber}
\tabnam{\pndata}
\def\tableunderline{\noalign{\vskip0.1truecm}\noalign{\vskip0.1truecm}}
\caption{{\bf Table \pndata.} Data on the radial velocities of the halo
planetary nebulae of M31. Listed are Galactic coordinates ($\ell,b$),
observed heliocentric radial velocities ($v_\odot$) in kms$^{-1}$
together with projected distances $R$ from the centre of M31 in kpc
and corrected heliocentric radial velocities ($\vlosc$) in \kms.  See
Section~5.1 for a discussion of the velocity corrections. Source:
Nolthenius \& Ford (1987)}
\halign{#\hfil&\quad\hfil#\hfil&\quad\hfil#\hfil&\quad\hfil#\hfil&\quad\hfil#\hfil&\quad\hfil#\hfil\cr
\noalign{\hrule}
\noalign{\vskip0.1truecm}
Name&$\ell$&    $b$& $\vlos$ & $R$ & $\vlosc$\cr
\tableunderline
SW3/A1 & 120.9 & -22.9 & -407  $\pm$ 10& 18.4 & -108\cr \tableunderline
SW3/A2 & 120.7 & -22.9 & -444 $\pm$ 10&  18.4 & -144\cr \tableunderline
NE6/8 & 121.8 & -20.2 & -79 $\pm$ 10& 19.7 & 222\cr \tableunderline
SW4/6 & 119.8 & -22.3 & -263 $\pm$ 10& 19.8  & 40\cr \tableunderline
SW4/5 & 119.7 & -22.3 & -511 $\pm$ 10& 20.0  & -208\cr \tableunderline
SW3/1 &  120.8 &  -23.4 & -263 $\pm$ 10& 24.4 & 36\cr \tableunderline
SW4/2 & 119.3 & -22.3 & -349 $\pm$ 10& 24.9  & -46\cr \tableunderline
NE6/6 & 122.0 & -19.9 & -452 $\pm$ 10& 25.4 & -151\cr \tableunderline
NE1-2/2 & 123.0 & -19.8 & -117 $\pm$ 10& 33.3 & 182\cr \tableunderline
\noalign{\vskip0.1truecm}\noalign{\hrule}
}
\tabnew
\endtable
\beginfigure{\fignumber}
\fignam{\figcumul}
\centerline{\psfig{figure=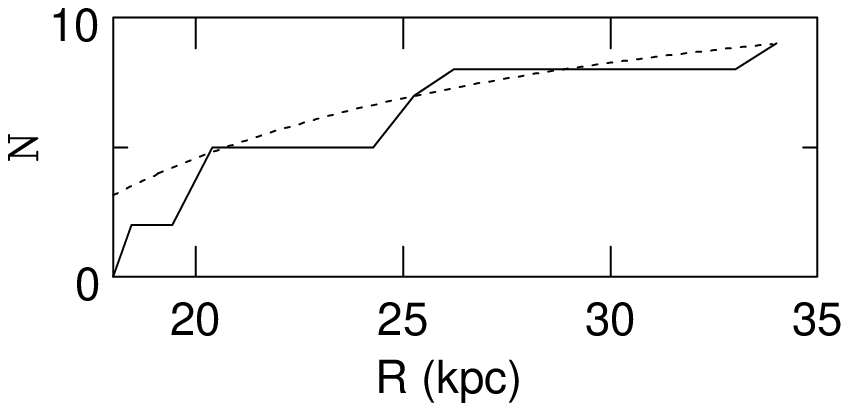,width=\hssize}}
\smallskip\noindent
\caption{{\bf Figure \figcumul.} Cumulative number plot for 
the satellite galaxies. Also shown is a dotted curve representing the
distribution from the assumed model, namely eq.~(12) with $\as = 250$
kpc.}
\fignew
\endfigure
\beginfigure{\fignumber}
\fignam{\figgcmodel}
\centerline{\psfig{figure=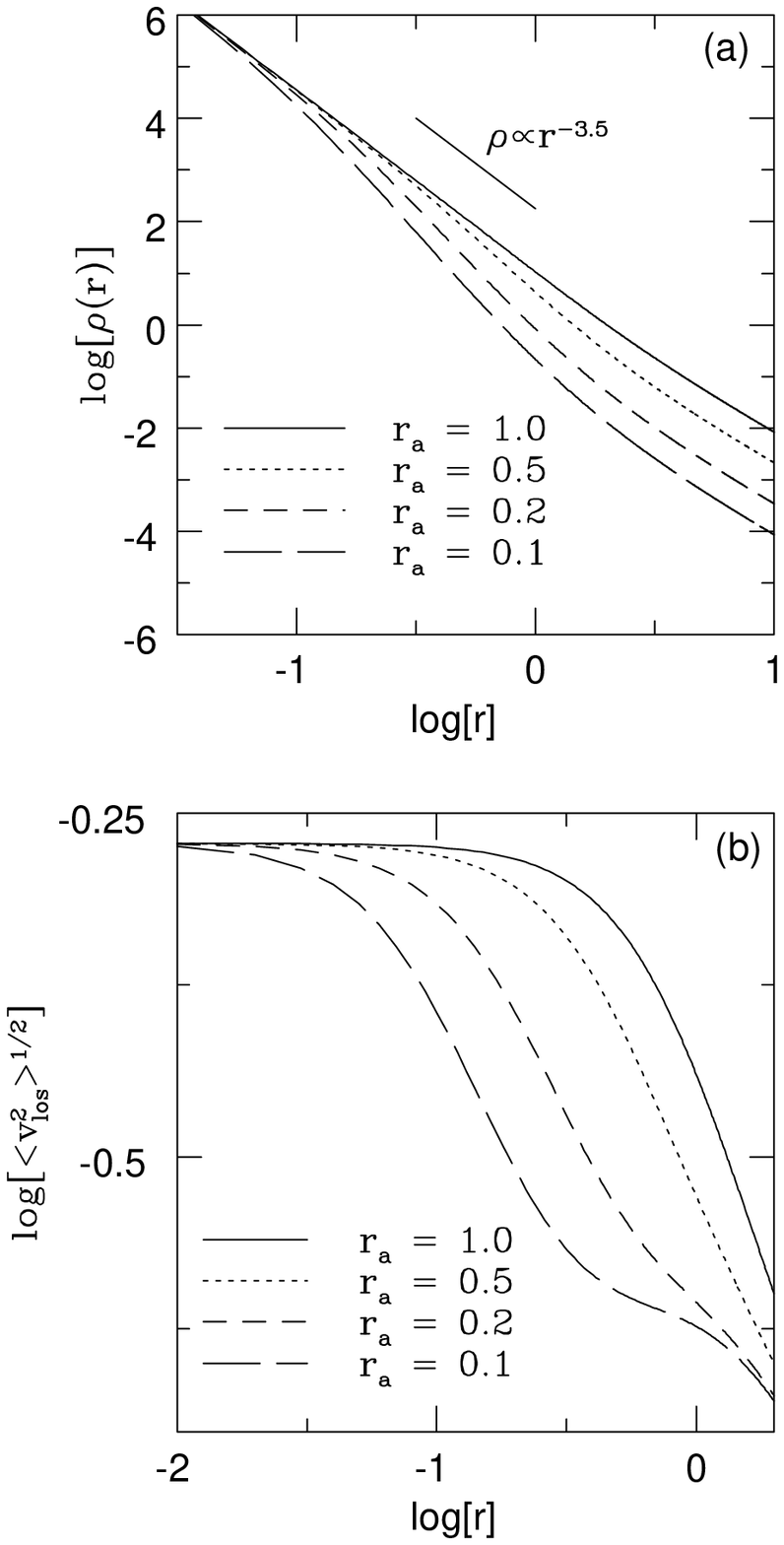,width=\hssize}}
\smallskip\noindent
\caption{{\bf Figure \figgcmodel.}(a) Radial profile for the globular
clusters corresponding to a density law of the form (17) with
$n=3.5$. Also shown is the slope of a simple power-law density of
index $3.5$. (b) Projected line of sight velocity dispersions for the
same model profiles. Units with $G=M=a=1$.}
\fignew
\endfigure
\section{The Data}
\noindent
In this section, we present the observational data which we use to
estimate the total mass. Data are available for three halo tracer
populations, namely the dwarf satellite galaxies, the outer globular
clusters and the halo planetary nebulae. Each of these populations has
different kinematics and so requires a different DF.
\subsection{The Satellites}
Table~\satdata\ presents the available data on the satellite galaxies
of M31 which have published radial velocity measurements. A number of
recent papers have contained similar tables (Mateo 1998; Lynden-Bell
1999; Courteau \& van den Bergh 1999; van den Bergh 1999b). These
authors all agree that the first eight satellites in our table are
definite members of the M31 subgroup of Local Group galaxies. We
follow Courteau \& van den Bergh (1999) in omitting both WLM (included
by Lynden-Bell) and EGB0247+63 (included by both Mateo and
Lynden-Bell) from our list.  These exclusions are made on the basis of
distance estimates -- the current distance estimate for WLM (Lee,
Freedman \& Madore 1993) locates it more than $800$ kpc from M31,
while the current distance estimate of 2.2 Mpc for EGB0247+63
(Karachentsev, Tikhonov \& Sazonova 1994) places it well outside the
Local Group.  Since Pegasus and IC 1613 both lie closer to M31 than
they do to the Milky Way, we have included them in our sample,
following Mateo (1998) and Lynden-Bell (1999). This is also suggested
by the three-dimensional diagram of the Local Group provided by Grebel
(1999).  However, van den Bergh (1999a) excludes both these satellites
from his list of the M31 subgroup.  One other possible uncertainty in
the dataset is the fact that the galaxies NGC 147 and NGC 185 may form
a binary system (van den Bergh 1998).  The data presented recently in
the literature are in good agreement with each other, except for the
distance estimates for the satellites. For this reason, we have
included in Table~\satdata\ the distances from Mateo (1998) and
Courteau \& van den Bergh (1999) to give an indication of the true
uncertainties.

To model the data in Table~1, let us assume that the satellite
distribution is spherically symmetric about the centre of M31 and that
the density $\rhos$ is a TF profile with scale-length $\as$
\eqnam{\satdens}
$$\rhos \propto {\as^2 \over r^2 (r^2 + \as^2)^{3/2}}.\eqno\new$$
Fig.~\figcumul\ shows how the cumulative number of satellites varies
with distance from M31. The figure also shows the profile
corresponding to a TF density model with $\as = 250$ kpc (dashed
curve) which we adopt as our standard model of the satellite number
density profile. We show later that our mass estimates are not overly
sensitive to the choice of this scale-length. DFs for the satellites
can be constructed both of constant anisotropy (\henonDF) and of
Osipkov-Merritt form (\anisoDF).
\subsection{The Globular Clusters}
M31 has approximately 300-400 globular clusters (Hodge 1992, Fusi
Pecci et al. 1993). The system comprises a rotating disk of metal-rich
clusters surrounded by a spherically symmetric, slowly rotating
distribution of metal-poor clusters (Elson \& Walterbos 1988). The
latter are of most interest to us here, as they probe the gravity
field at large distances.  Table~\gcdata\ presents the published data
on the seventeen globular clusters which lie at distances exceeding
$\sim 20$ kpc in projection from the centre of M31 and for which
radial velocities are available.

To model the globular cluster data, let us extend the idea of
isothermal populations (e.g., Binney \& Tremaine 1987; Evans 1993) to
the TF galaxy model.  Let us consider a DF of the form
\eqnam{\globDF}
$$ f(Q) = \rho_0 \left( { n \over 2 \pi v_0^2 } \right)^{3/2}
\exp\left({n Q \over v_0^2}\right),\eqno\new$$
where $\rho_0$ is a normalisation factor and $n$ is chosen to fit the
number density profile.  Crampton et al. (1985) analysed a large
sample of M31 globular clusters and argued that the radial profile is
similar to that of the Milky Way and so declines like $r^{-3.5}$ at
large radii (e.g., Djorgovski \& Meylan 1994; Ashman \& Zepf 1998).
Pursuing this analogy, we choose the parameter $n$, which controls the
radial fall-off, to be $3.5$.

As it stands, the DF (\globDF) describes a population with no net
rotation. The outer globular cluster system of M31 rotates with a
velocity of $\sim 80$ \kms\ (Huchra, Stauffer \& van Speybroeck
1982). So, we build a rotating DF by
\eqnam{\frot}
$$ f_{{\rm gc}}(Q) = T_{{\rm gc}} f_{{\rm even}}(Q) + 
     (1-T_{{\rm gc}})f_{{\rm odd}}(Q),\eqno\new$$
where $T_{{\rm gc}}$ is a parameter that controls the streaming.
The even and odd parts of the DF are given by
\eqnam{\feven}
$$ f_{{\rm even}}(Q) = f(Q),\eqno\new$$
and
\eqnam{\fodd}
$$f_{{\rm odd}}(Q) = \cases{2f(Q), & if $v_{\phi} \geq 0$,\cr 0,&
otherwise.\cr} \eqno\new$$
Here, $v_\phi$ is the azimuthal streaming velocity, where the rotation
axis is taken as perpendicular to the disk of M31. The inclination of
the disk of M31 is $77.5^\circ$ (Hodge 1992).  Choosing $T_{{\rm
gc}} = 0.3$ gives the required degree of rotation ($\sim 80$ \kms) of
the system at the radii at which we possess data ($R \sim 20-30 $
kpc).

If we assume a TF model for the halo, the expressions for the density
law $\rho_{\rm gc}$ and velocity dispersions $\langle v_r^2 \rangle$
corresponding to the DF (\globDF) are quite complicated (see Wilkinson
1999). However, they can be greatly simplified when $r < \ra$ and $r <
a$. This is the relevant limit, as the globular cluster data are
confined to the inner parts of the halo. In this case, the density
profile reduces to a simple power-law with index $n$, namely
\eqnam{\globdens}
$$\rho_{\rm gc}(r) \sim \rho_0\left({2a \over r}\right)^n,\eqno\new$$
while the radial velocity dispersion is just
$$\langle v^2_{r}\rangle \sim {v_0^2 \over n}.\eqno\new$$
The reason for the elegant limits is that the inner parts of the TF
model look like an isothermal sphere, and so the DF (\globDF)
generates a population with a power-law density profile and a constant
velocity dispersion. Fig.~\figgcmodel\ shows the density profile and
projected line of sight velocity dispersions for this model for a
range of values of $\ra$, from which the isothermal limit can be
distinguished.  
\subsection{The Planetary Nebulae}
Planetary nebulae are luminous enough to be detected in external
galaxies, and they have been increasingly exploited in recent years to
estimate enclosed masses (e.g., Bridges 1999; Arnaboldi, Napolitano \&
Capaccioli 2000). Over 500 planetary nebulae have been detected in
M31, although the total population may exceed $\sim 10^4$ (Hodge
1992). Many of these reside in the inner bulge and disk. For the outer
disk and halo, the largest currently available dataset is that of
Nolthenius \& Ford (1987). They obtained radial velocities of 34
planetary nebulae and classified them into halo and disk objects. This
classification is not straightforward as it depends on the details of
the adopted warped disk model. Nonetheless, approximately 9 of the
planetary nebulae are residents of the halo and they are listed in
Table~\pndata . The planetary nebulae system rotates with a velocity
of $92 \pm 43$ \kms\ in the same sense as the rotation of the disk
(Nolthenius \& Ford 1987).

We model the system in a similar way to the globular clusters. We
assume a DF of the form
\eqnam{\f_rotpn}
$$ f_{{\rm pn}}(Q) = T_{{\rm pn}} f_{{\rm even}}(Q) +
(1-T_{{\rm pn}})f_{{\rm odd}}(Q),\eqno\new$$
implying a density law (\globdens).  In order to fit the number
density of the planetary nebulae, we choose $n=4.8$ in the density
law, as suggested by Nolthenius \& Ford's (1987) fit.  We pick
$T_{{\rm pn}} = 0.065$ to ensure that streaming velocity $\langle
v_{\phi}\rangle \sim 92$ \kms. Let us note that the value of $T$ is
quite different to that obtained for the globular clusters, although
the rotation velocities are comparable. The reason for this is that
$T$ depends on the choice of asymptotic fall-off $n$ in the density
law.
\section{The Algorithm}
We derive our probability formulae using only the assumption
that the halo potential of Andromeda is spherically symmetric. We seek
to maximise the likelihood of a parameterised model with respect to
the data $P(\model|\data)$. This is given by Bayes' theorem as (see
Little \& Tremaine 1987; WE99):
\eqnam{\modelprob}
$$P(\model|\,\data) P(\data) = P(\model) P(\data|\,\model).\eqno\new$$
Here, $P(\model)$ describes our prior beliefs as to the likelihood of
the model parameters, while $P(\data|\,\model)$ is the probability of
the data given the model. For each of the $S$ satellites of M31, the
data consist of their three-dimensional positions $r_i$ ($i = 1\dots
S$) with respect to the centre of Andromeda, together with their
heliocentric line of sight velocities $\vlosi$. For the globular
clusters and planetary nebulae, the three-dimensional positions are
unknown. So, for each of the $G$ globular clusters and $N$ planetary
nebulae, the data are the two-dimensional projected positions $R_i$
($i = 1\dots G+N$) with respect to the centre of Andromeda, together
with their line of sight velocities $\vlosi$.
\beginfigure{\fignumber}
\fignam{\figangles}
\centerline{\psfig{figure=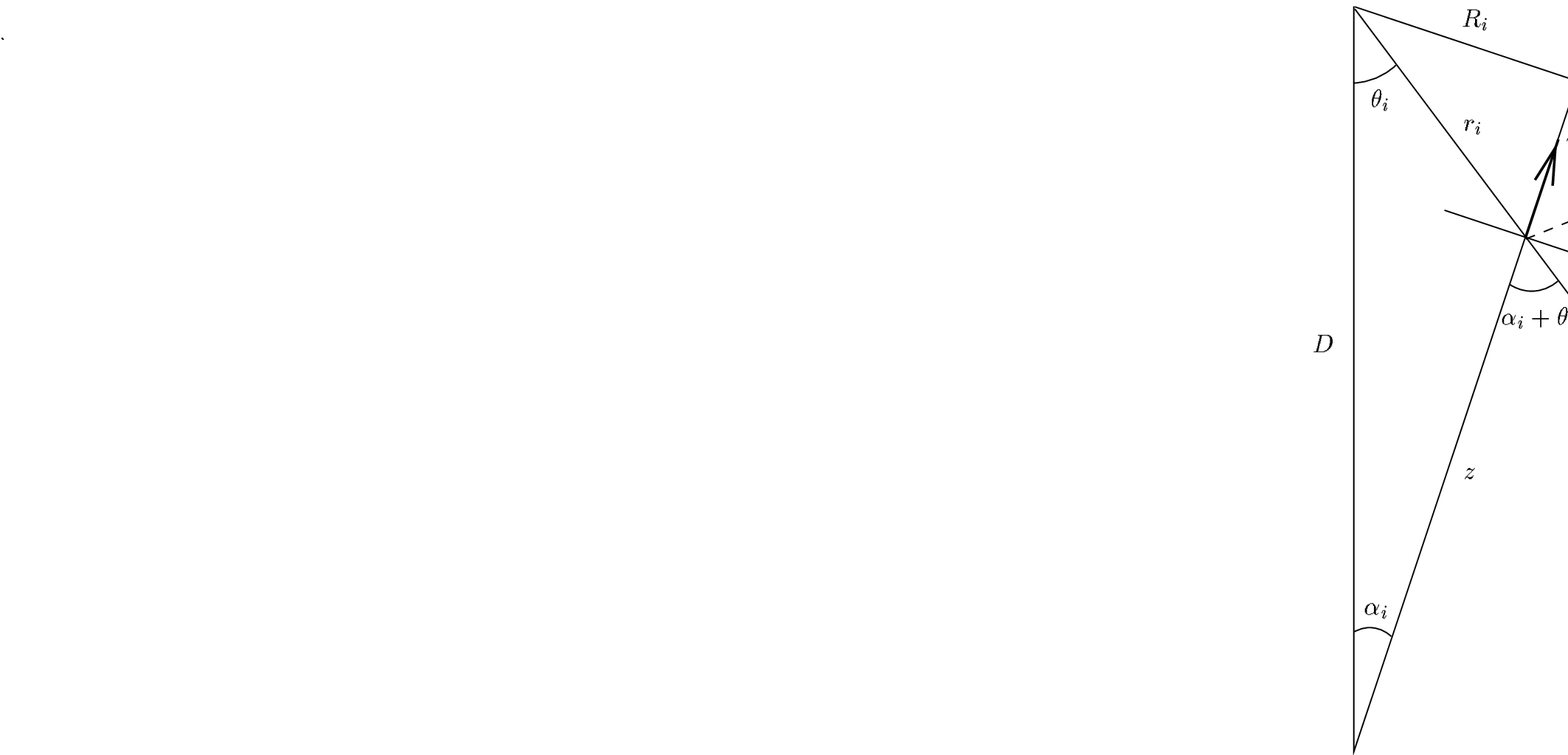,height=\hssize}}
\smallskip\noindent
\caption{{\bf Figure \figangles.} This figure shows a satellite at true
position $r_i$ and projected position $R_i$ with respect to the centre of
the spherical halo of the Andromeda galaxy. The angles $\theta_i$
and $\alpha_i$ are defined. The index $i$ labels the number of 
satellites.}
\fignew
\endfigure
\subsection{Probabilities for the Satellites}
Let $\vlosic$ be the line of sight velocity corrected for the solar
motion and the radial motion of M31. Then, the probability of finding
a satellite at position $r_i$ moving with line of sight velocity
$\vlosic$ is simply
\eqnam{\simpleprob}
$$P(r_i,\vlosic|\model) = {1 \over \rhos}\int d^3v\,f(\varepsilon,l)
\delta(\vlosc - \vlosic),\eqno\new$$
where $\rhos$ is the density distribution of the satellites and
$f(\varepsilon,l)$ is the DF. As some of the satellites of Andromeda
lie at large angular separations from its centre, it is important to
carry out the integration over the velocities perpendicular to the
line of sight from the perspective of an observer at a finite distance
$D = 770$ kpc and not at infinity. Let the components of the
satellite's velocity with respect to a spherical polar coordinate
system oriented at Andromeda's centre be ($v_r, v_\theta,
v_\phi$). Along any line of sight, let ($\vT, \eta$) be polar
coordinates in the plane of projection and $\vlosc$ the component
along the line of sight and perpendicular to the plane of projection.
Fig.~\figangles\ shows a satellite galaxy at a spherical polar
distance $r_i$ and a projected distance $R_i$. The line of sight
subtends an angle $\alpha_i$ from the axis joining the observer to the
centre of Andromeda, while the radius vector subtends an angle
$\theta_i$. It is straightforward to see:
\eqnam{\defangles}
$$\sin (\alpha_i + \theta_i) = R_i/r_i, \qquad\qquad \sin \alpha_i =
R_i/D.\eqno\new$$
The angles $\alpha_i$ and $\theta_i$ are fixed for the $i$th satellite
galaxy since we know its distance along the line of sight. The
relations between the observed velocity and the velocity components in
spherical coordinates centred on M31 are
\eqnam{\transform}
$$\eqalign{v_r = &\vT \cos \eta \sin (\theta_i + \alpha_i) - 
\vlosic \cos (\theta_i + \alpha_i),\cr
v_\theta^2 + v_\phi^2 = &[\vlosic \sin(\theta_i + \alpha_i) + 
\vT \cos \eta \cos (\theta_i + \alpha_i)]^2 \cr
&\qquad\qquad\qquad + \vT^2 \sin \eta^2.\cr}\eqno\new$$
Thus the required probability is given by
\eqnam{\satprob}
$$\eqalign{P(r_i,\vlosic|\model) = {1 \over
\rhos(r_i)}&\int_0^{\sqrt{2\psi(r_i)-\vlosic^2}}
dv_t\, v_t\cr &\quad\times \int_0^{2\pi}d\eta\, f(\varepsilon,l)}
\eqno\new$$
Note that the transformations (\transform) are used to rewrite the
binding energy $\varepsilon$ and angular momentum per unit mass $l$ in
terms of the integration variables. This formula is valid for any
spherically symmetric density distribution of satellite galaxies
$\rhos$. If the observer is infinitely far away from the galaxy, then
all the above formulae hold good with $\alpha_i = 0$.
\subsection{Probabilities for the Globular Clusters and Planetary Nebulae}
The probability of finding a globular cluster at projected radius
$R_i$ moving with line of sight velocity $\vlosic$ is more cumbersome
as we must integrate along the line of sight. As a result, the angle
$\theta$ is no longer fixed for each object but is a function of the
distance along the line of sight $z$. The probability of drawing the
data pair ($R_i, \vlosic$) from the projected distributions is
\eqnam{\projprob}
$$\eqalign{P(R_i,\vlosic|\model) = {1 \over \Sigma(R_i)}\int_{-\infty}^\infty
dz  &\int d^3v\,  f(\varepsilon, l)\cr
&\times \delta(\vlosc - \vlosic).\cr}\eqno\new$$
For an observer at finite distance $D$, we then obtain
\eqnam{\projprob}
$$\eqalign{P(R_i,\vlosic|&\model) = {R_i \over \Sigma (R_i)}
\int_{\theta_{{\rm min}}}^{\theta_{{\rm max}}} d\theta\,
\cosec^2(\alpha_i+\theta)\cr &\times
\int_0^{\sqrt{2\psi(\theta)-\vlosic^2}}
dv_t\,v_t\int_0^{2\pi}
d\eta f(\varepsilon,l),\cr}\eqno\new$$
where
\eqnam{\thetamin}
$$\theta_{{\rm min}} = \cases{\acos(R_i/D) - \acos(R_i/r_{{\rm
max}}), & if $\rmax < D$,\cr 0, &if $\rmax >
D,$}\eqno\new$$
and
\eqnam{\thetamax}
$$\theta_{{\rm max}} = \acos(R_i/D) + \acos(R_i/r_{{\rm max}}).\eqno\new$$
In all the above formulae, $\rmax$ is the solution of the equation
$\psi(\rmax) = \onetwo \vlosic^2$, namely
$$\rmax = a \csch \Bigl( {\vlosic^2 \over 2 v_0^2} \Bigr).$$
Physically, $\rmax$ is the greatest radius at which the globular
cluster or planetary nebula could have the line of sight velocity
$\vlosic$ and still remain bound to the Andromeda galaxy.

We note that for an observer who is infinitely far away, all the above
formulae hold good with
\eqnam{\infthetamin}
$$\theta_{{\rm min}} = \pi/2 - \acos(R_i/r_{{\rm max}}),\eqno\new$$
and
\eqnam{\infthetamax}
$$\theta_{{\rm max}} = \pi/2 + \acos(R_i/r_{{\rm max}}).\eqno\new$$
It is often possible to carry out the velocity integration in
(\projprob) analytically leaving only a double integral to be
performed numerically.
\beginfigure{\fignumber}
\fignam{\figL2BCont}
\centerline{\psfig{figure=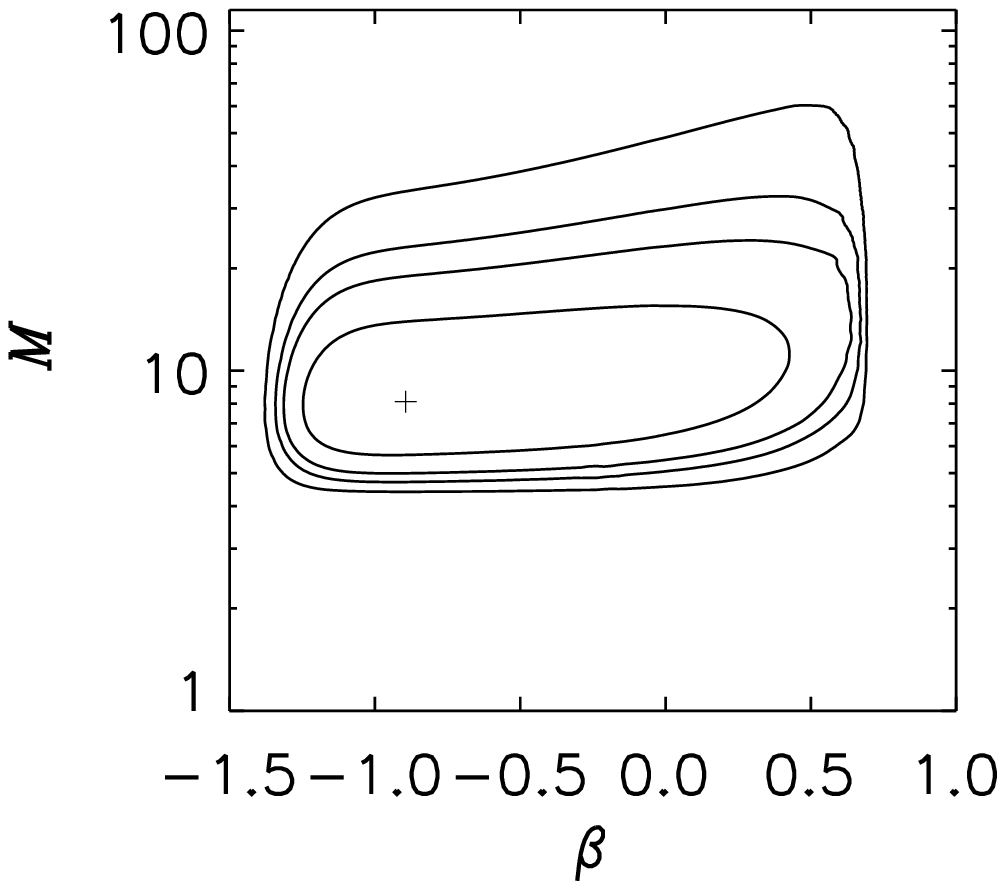,width=\hssize}}
\smallskip\noindent
\caption{{\bf Figure \figL2BCont.} Likelihood contours for the total mass $M$
(in units of $10^{11}\Msun$) and the anisotropy parameter $\beta$,
assuming a DF of the form (\henonDF). Contours are at heights of 0.32,
0.1, 0.045 and 0.01 of peak height. Data on the satellite galaxies
only have been used.}
\fignew
\endfigure
\subsection{Prior Probabilities and Error Convolution}
In any Bayesian analysis, it is necessary to specify the prior
probabilities of the model parameters.  We have meagre information
about the anisotropy of the velocity distribution. When using constant
anisotropy DFs (\dejonghe), we choose $P(\beta) \propto
1/(3-2\beta)^2$, as this ensures that the ratio of radial kinetic
energy to total kinetic energy is uniform. When using DFs of
Osipkov-Merritt form (\anisoDF), we choose a prior of $1/\ra$ for the
anisotropy radius.  This is the recommended choice of an unbiased
prior for a parameter which can vary in the range ($0,\infty$) (see
e.g., Kendall \& Stuart 1977). In our earlier analysis of the mass of
the Milky Way (WE99), we experimented with a prior on $M$ of $1/M^2$
and found it yielded good results.  This has some advantages over
$1/M$, as it reduces the probability of unrealistically large halos.
Since it is important to verify that our results are not overly
sensitive to the choice of priors, Section~5 also presents results
using a number of alternatives.

The projected positions of the globular clusters and planetary nebulae
are known to good accuracy, as are the radial velocities of all the
objects in our tracer populations. However, the distances to the
satellite galaxies are uncertain. We take this into account with an
error convolution function to obtain the probabilities of the
observations from our theoretical probabilities.  As there is no
reason to assume that the observational errors are Gaussian, we use
hat-box functions. These have the form
\eqnam{\hatbox}
$$ B(d;d_{\rm min},d_{\rm max}) = \cases{ \displaystyle{ 1\over 
d_{\rm max} - d_{\rm min}},  & $d_{\rm min} < d
< d_{\max}$,\cr 
\null&\null \cr
\displaystyle{0}, & otherwise.}\eqno\new$$
Our error convolution aims to take account of two
factors. First, there is the quoted error estimate associated
with each distance given in Table~\satdata. Second, there is a small,
but distinct, probability $\epsilon$ that some of the published
distance estimates are seriously in error due to systematic
uncertainties. Therefore, we choose our convolution function $E$ to
have the form (c.f., Schmoldt \& Saha 1998)
\eqnam{\hatbox}
$$\eqalign{E(z;i) = (1-\epsilon) & B(z;s_i-\Delta s_i,s_i+\Delta
s_i)\cr & + \epsilon B(z;s_{\rm min},s_{\rm max}),}\eqno\new$$
where $s_i$ is the published distance estimate of the $i$th satellite
with error $\Delta s_i$ and $s_{\rm min}$ and $s_{\rm max}$ are the
minimum and maximum of all the published distance estimates listed in
Table~\satdata. In what follows, we assume that the probability of a
rogue distance $\epsilon$ is $0.1$.  Section~5 presents results both
for the case in which $\Delta s_i$ is taken to be the quoted error on
the distance measurement given in Table~\satdata\ and for the worst
case in which the distance errors are assumed to be $25\%$. This value
is chosen based on the magnitudes of recent revisions of the distance
estimates (see columns~5 and~6 of Table~\satdata).
\section{Mass Estimates}

\noindent
In this section, we apply the algorithm to the observational data.
Section 5.1 considers only the satellite galaxies. Then, in Section
5.2, each of our three tracer populations is analysed separately
before they are combined to form a single sample.  The robustness of
our estimates is examined in Section 5.3

\subsection{Satellite Galaxies Only}
\noindent
First, we must convert the observed heliocentric line of sight
velocities $\vlos$ into velocities in the rest frame of Andromeda
$\vlosc$. This entails removing the contributions from both the solar
motion within the Galaxy and the Galactic motion towards M31.  To do
this, we assume a circular speed of 220 kms$^{-1}$ at the
Galactocentric radius of the sun ($R_{\odot}$ = 8.0 kpc) and a solar
peculiar velocity of ($U,V,W$) = (-9,12,7), where $U$ is directed
outward from the Galactic Centre, $V$ is positive in the direction of
Galactic rotation at the position of the sun, and $W$ is positive
towards the North Galactic Pole. The line of sight velocity of M31,
corrected for the motion of the sun, is $v_{r,{\rm M31}} = -123$ \kms,
based on the uncorrected value of $v_{\odot,{\rm M31}} = -301$ \kms\
(Courteau \& van den Bergh 1999).  Some of the satellites are located
at large angular distances from the centre of M31. This means that the
unknown tangential velocity of M31 contaminates the observed
heliocentric radial velocity. In the absence of any measurement, we
assume that this unknown tangential velocity is zero. This is very
reasonable, as there are no other large galaxies in the Local Group to
generate angular momentum through tidal torques, and so -- as Kahn \&
Woltjer (1959) originally argued -- we are probably falling directly
towards Andromeda. In order to correct for the motion towards M31, we
therefore subtract the component of $v_{r,{\rm M31}}$ along the line
of sight to each satellite.  As pointed out by Bahcall \& Tremaine
(1981), the large angular separations from M31 of both IC 1613 and
Pegasus mean that the determination of their velocities relative to
M31 depends sensitively on any tangential velocity component of M31.

As a first calculation, we analyse the satellite dataset using the
constant anisotropy family of DFs of the form (\henonDF). Due to the
singularity in the DF that can occur at the radial orbits ($l=0$),
some care is needed when carrying out the integration in (\satprob)
over the transverse velocities -- the relevant substitutions are given
in Appendix~A.  Fig.~\figL2BCont\ presents likelihood contours in the
$M$-$\beta$ plane for the satellite dataset only. The best estimate
for the total mass is $8.1 \times 10^{11}\Msun$. The most likely
value of $\beta$ is $-0.9$, indicating a somewhat tangential velocity
distribution. We note that the likelihood contours are elongated in
the direction of the $\beta$-axis, indicating that the anisotropy is
more poorly constrained than the mass.  For comparison, our earlier
estimate for the Milky Way halo (WE99) was $19 \times
10^{11}\Msun$. We reach the surprising conclusion that {\it the
total mass of M31 may be less than that of the Milky Way}.

There is a second piece of evidence worth bringing forward.  For an
isotropic tracer population falling off like $r^{-3}$ in a galaxy with
a flat rotation curve of amplitude $v_0$, there is the simple result
(e.g., Evans, H\"afner \& de Zeeuw 1997)
$$v_0^2 = 3 \langle v_r^2 \rangle.\eqno\new$$
Here, $\langle v_r^2 \rangle$ is the mean square radial velocity of
the sample (in the rest frame of the galaxy).  Applying this formula
to the dataset of the Milky Way satellites gives $v_0 \sim 220$
\kms. This is in excellent agreement with estimates of the local
circular speed, from which we conclude that the Milky Way's halo is
roughly isothermal out to the distances probed by the satellites. The
dataset on the Andromeda satellites gives $v_0 \sim 140$ \kms, which
is substantially less than amplitude of the HI rotation curve ($\sim
240$ \kms).  This suggests that the isothermal region of the Andromeda
halo is less than the volume sampled by the satellite galaxies.

\beginfigure*{\fignumber}
\fignam{\figcontours}
\centerline{\psfig{figure=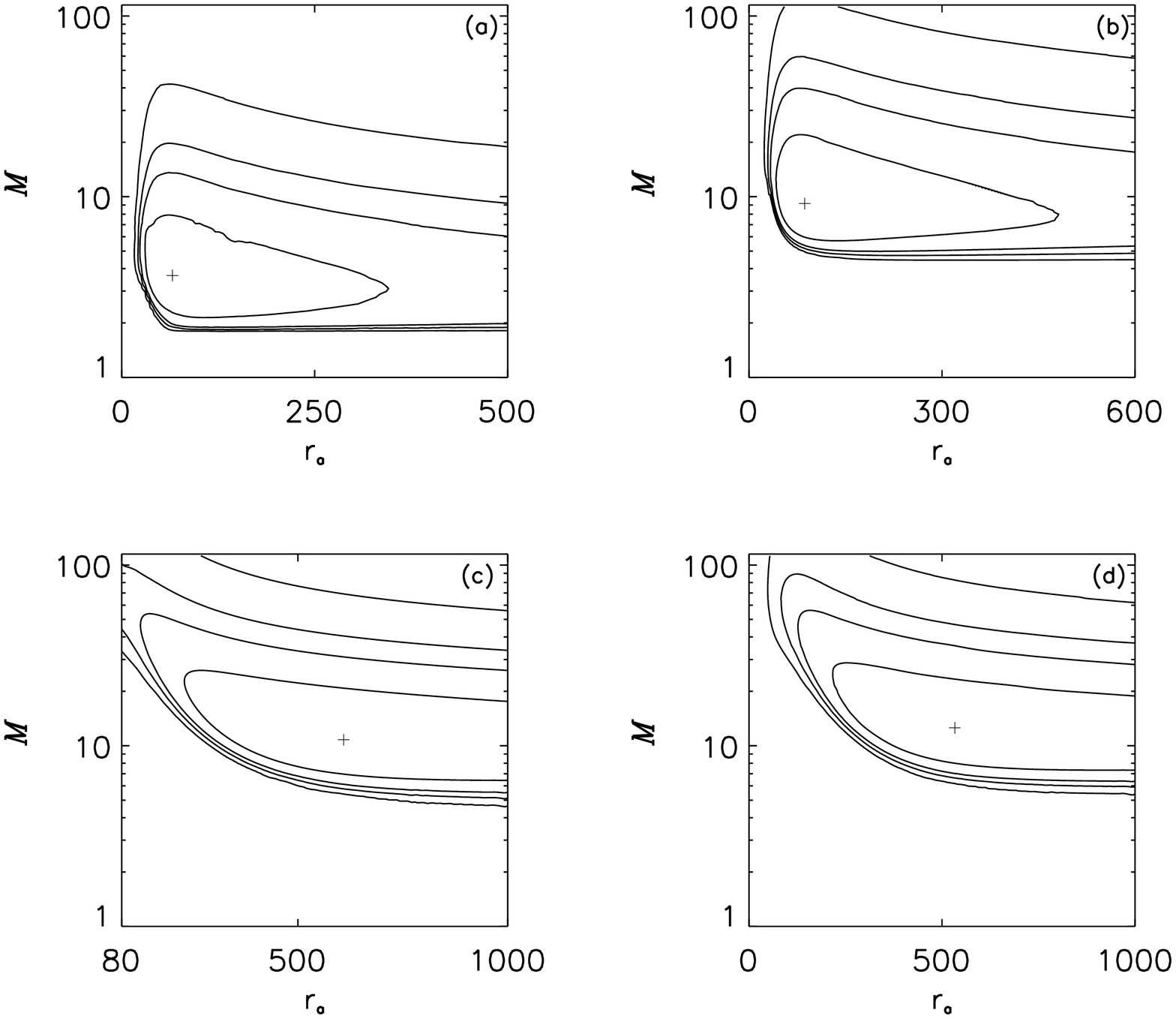,width=\hsize}}
\smallskip\noindent
\caption{{\bf Figure \figcontours.} Likelihood contours for the total mass $M$
(in units of $10^{11}\Msun$) and the anisotropy radius $\ra$ (in
kpc). Contours are at heights of 0.32, 0.1, 0.045 and 0.01 of peak
height. Data used: (a) Planetary nebulae only, (b) Globular clusters
only, (c) Satellite galaxies only, (d) All data combined.}
\fignew
\endfigure
\begintable*{\tabnumber}
\tabnam{\resultsa}
\def\tableunderline{\noalign{\vskip0.1truecm}\noalign{\vskip0.1truecm}}
\caption{{\bf Table \resultsa.} Most likely values of $M$ (in units of 
$10^{11}\Msun$) and $\ra$ (in kpc) obtained from the contour plots
of Fig.~\figcontours. The corresponding values of a (in kpc) and the
mass inside 30 kpc are given. The effects of changing the prior on
$\ra$ and $M$ are illustrated.}
\halign{#\hfil&\quad\hfil#\hfil&\quad\hfil#\hfil&\quad\hfil#\hfil&\quad\hfil#\hfil&\quad\hfil#\hfil&\quad\hfil#\hfil\cr
\noalign{\hrule}
\noalign{\vskip0.1truecm}
Data& $M$ prior & $\ra$ prior & Most likely & Most likely & Most
likely & M ( $< 30$ )\cr
& & & $\ra$ & $M$ & $a$ &\cr \tableunderline
PN & $1/M^2$ & $1/\ra$ & 66 & 3.7 & 26 & 2.8 \cr \tableunderline
PN & $1/M$ & $1/\ra$ & 35 & 5.6 & 41 & 3.3 \cr \tableunderline
PN & $1/M^2$ & $1$ & 500 & 3.1 & 21 & 2.5 \cr \tableunderline
\null&\null &\null &\null & \null & \cr 
GC & $1/M^2$ & $1/\ra$ & 86 & 9.2 & 68 & 3.7 \cr \tableunderline
GC & $1/M$ & $1/\ra$ & 75 & 11.4 & 85 & 3.8 \cr \tableunderline
GC & $1/M^2$ & $1$ & 599 & 7.9 & 58 & 3.6 \cr \tableunderline
\null&\null &\null &\null & \null & \cr 
Sat & $1/M^2$ & $1/\ra$ & 610 & 10.8 & 80 & 3.8 \cr \tableunderline
Sat & $1/M$ & $1/\ra$ & 482 & 13.7 & 102 & 3.9 \cr \tableunderline
Sat & $1/M^2$ & $1$ & 1000 & 9.7 & 72 & 3.7\cr \tableunderline
\null&\null &\null &\null & \null & \cr 
All & $1/M^2$ & $1/\ra$ & 551 & 12.3 & 91 & 3.8 \cr \tableunderline
All & $1/M$ & $1/\ra$ & 439 & 15.5 & 115 & 3.9 \cr \tableunderline
All & $1/M^2$ & $1$ & 1000 & 10.6  & 79 & 3.8 \cr \tableunderline
All & $1/M$ & $1$ & 1000 & 12.6 & 95 & 3.9 \cr \tableunderline
\noalign{\vskip0.1truecm}\noalign{\hrule}
}
\tabnew
\endtable
\subsection{All Data}
\noindent
We now proceed to subject our hypothesis to serious scrutiny by
modelling the data not just on the satellite galaxies, but also the
planetary nebulae and the globular clusters as well. For all the
samples, we adopt the alternative family of DFs, namely those of
Osipkov-Merritt form (\anisoDF), as a precautionary check for
robustness.

Fig.~\figcontours(a) shows likelihood contours in the plane of the
total mass $M$ and the anisotropy radius $\ra$ for the planetary
nebulae. This dataset implies a low halo mass of $\sim 3.7 \times
10^{11}\Msun$ and an anisotropy radius of $\sim 66$ kpc. Given
that the planetary nebulae are all located at projected radii between
$18$ and $34$ kpc, this mass estimate must be taken as a constraint
only on the mass inside a three dimensional radius of $\sim 31$ kpc.
This figure is obtained by taking the median projected radius and
converting to a three dimensional radius by multiplying by a
deprojection factor of $\pi /2$. It is worth noting that the available
data on the planetary nebulae are not uniformly distributed over the
halo of M31, but are concentrated in two regions near the optical
disk. A total mass of $3.7 \times 10^{11}\Msun$ implies that the
mass within $\sim 31$ kpc is $2.8\times 10^{11}\Msun$.

Fig.~\figcontours(b) shows the likelihood contours obtained from the
globular cluster dataset. In this case, the mass estimate is $\sim 9.2
\times 10^{11}\Msun$ with an anisotropy radius of $\sim 86$
kpc. The globular cluster data probe a range of projected radii
between $19$ and $34$ kpc. They are more uniformly distributed than
the planetary nebulae making this estimate more robust. A total halo
mass of $9.2 \times 10^{11}\Msun$ implies a mass inside $ 40$ kpc
of $4.7 \times 10^{11}\Msun$. We note that the globular clusters
are effectively probing a volume which is approximately double that
probed by the planetary nebulae.  The globular cluster data imply a
mass inside $31$ kpc of $3.8\times 10^{11}\Msun$, which is
somewhat larger than that implied by the planetary nebulae.

Fig.~\figcontours(c) presents the results obtained using the data on
the satellite galaxies. In this case, the best mass estimate is $\sim
10.8 \times 10^{11}\Msun$ with an anisotropy radius of $\sim 610$
kpc. The large value of the anisotropy radius suggests that the
velocity distribution is isotropic over the entire region probed by
the satellites. Given that the satellites probe radii from $\sim 10$
kpc to $\sim 500$ kpc, this is an estimate of the total mass of the
M31 halo.  In Fig.~\figcontours(d), all the data from our three
populations are combined into a single dataset. The mass estimate is
$\sim 12.3 \times 10^{11}\Msun$ with an anisotropy radius of $\sim
551$ kpc. This corresponds to a halo scalelength $a \sim 91$ kpc. The
value of the total mass implies masses interior to $31$ kpc and $40$
kpc of $\sim 4.0 \times 10^{11}\Msun$ and $\sim 4.9 \times
10^{11}\Msun$, respectively. These values are in good agreement with
the results obtained from the globular cluster data. The mass inside
$31$ kpc is greater than that implied by the planetary nebulae data --
the non-uniform distribution of the planetary nebulae is the most
likely cause of this discrepancy. Our final calculation assumes that
all three populations have the same anisotropy radius; this seems
reasonable as the individual calculations in Fig.~\figcontours(a)-(c)
all yield values of $\ra$ greater than the outermost datapoints,
implying the respective velocity distributions are all largely
isotropic.

Table~\resultsa\ summarises the results described above and
illustrates the effects of varying the prior probabilities. For the
planetary nebulae data, changing the priors has some effect on the
mass estimates obtained. Changing the prior on $M$ to $1/M$ allows
larger values of $M$ and the mass estimate rises by $\sim 50\%$. A
uniform prior on $\ra$ naturally allows larger values of $\ra$, but
changes the mass by only $\sim 16\%$. For the globular cluster data,
the situation is somewhat better, with the largest change being a
$24\%$ increase in the mass estimate when the prior on $M$ is changed
to $1/M$. For the complete dataset, allowing larger halos produces an
increase in the mass estimate of $\sim 26 \%$, which is a reasonably
small change. Indeed, if we simultaneously relax both the $M$ and
$\ra$ priors, the mass estimate changes by only $\sim 2\%$.  Given the
other uncertainties in the problem (see Section~6), sensitivity to
the priors is not a serious worry.

We remark that all the results in Table~\resultsa\ support the
hypothesis that the mass of the Andromeda halo is probably less than
that of the Milky Way halo (see Table 5 of WE99 as a comparison).
\begintable*{\tabnumber}
\tabnam{\resultsb}
\def\tableunderline{\noalign{\vskip0.1truecm}\noalign{\vskip0.1truecm}}
\caption{{\bf Table \resultsb.} Table illustrating the effects of
changing some of the assumed model parameters. The parameter which has
been changed is given in the third column. The prior on $M$ is
$1/M^2$, the prior on $\ra$ is $1/\ra$. Most likely values of $M$ (in
units of $10^{11}\Msun$) and $\ra$ (in kpc) are given together
with the corresponding values of a (in kpc) and the mass inside 30
kpc. Sources for data: 1. Courteau \& van den Bergh (1999) 2. Mateo
(1998)}
\halign{#\hfil&\quad\hfil#\hfil&\quad\hfil#\hfil&\quad\hfil#\hfil&\quad\hfil#\hfil&\quad\hfil#\hfil&\quad\hfil#\hfil\cr
\noalign{\hrule}
\noalign{\vskip0.1truecm}
Data& Source &  Comment & Most likely & Most likely & Most
likely & M ( $< 30$ )\cr
& & & $\ra$ & $M$ & $a$ &\cr \tableunderline
Sat & 1 & DF $\sim l^{-2\beta}f(\varepsilon)$ & $\beta = -0.9$ & 8.1 &
60 & 3.6\cr \tableunderline
Sat & 1 & Pegasus \& IC 1613 & 128 & 7.2 & 53 & 3.6 \cr 
& & omitted & & & &  \cr \tableunderline
All & 1 & Pegasus \& IC 1613 & 114 & 9.4 & 69 & 3.7 \cr 
& & omitted & & & &  \cr \tableunderline
All & 2 & Sat. distances & 607 & 13.2 & 98 & 3.9 \cr
 & & altered & & & & \cr \tableunderline
All & 1 & $D = 900$ kpc & 615 & 15.7 & 117 & 3.9 \cr \tableunderline
All & 1 & $\Delta s = 25\%$ & 535 & 12.6 & 94 & 3.8 \cr
\tableunderline
All & 1 & $\as = 150$ kpc & 551 & 13.0 & 97 & 3.9 \cr \tableunderline
All & 1 & $n_{\rm gc} = 5$ & 619 & 10.8 & 80 & 3.8 \cr \tableunderline
All & 1 & $T_{\rm gc} = 1$ & 537 & 12.6 & 94 & 3.8 \cr \tableunderline
All & 1 & $\vc = 270$\kms & 613 & 11.2 & 65 & 4.7 \cr 
& & $\vc = 200$\kms & 539 & 12.6 & 135 & 2.7\cr \tableunderline
\noalign{\vskip0.1truecm}\noalign{\hrule}
}
\tabnew
\endtable

\subsection{Robustness}
\noindent
We now look at the effects of altering some of the assumptions made in
the modelling. The calculations are summarised in Table~\resultsb.

First, it is important to assess whether the mass estimate is strongly
sensitive to any one of the datapoints, in a manner analogous to the
effect of Leo I on mass estimates of the Milky Way. We remove each of
the satellites in turn and apply the algorithm to the reduced
dataset. The only satellite which significantly alters the answer is
Pegasus -- removing Pegasus from the dataset, reduces the mass
estimate based only on the satellites by $\sim 23
\%$. Removing any of the other satellites affects the mass estimate by
less than $10\%$.  The dataset used by Courteau \& van den Bergh
(1999) excludes both Pegasus and IC 1613. With this subsample, we
obtain a low mass estimate of $7.2 \times 10^{11}\Msun$ (using our
standard $1/M^2$ and $1/\ra$ priors).  For comparison, Courteau \& van
den Bergh (1999) obtained $13.3 \pm 1.8 \times 10^{11}\Msun$ based on
the projected mass estimator of Heisler, Tremaine \& Bahcall (1985).

Up to this point, we have used the distances to the satellites in
Courteau \& van den Bergh (1999). If instead, we take the distances
from Mateo (1998), we find that our mass estimate based on the entire
dataset increases by $\sim 7\%$ to $13.2 \times 10^{11}\Msun$. Mateo's
distances imply larger separations from M31 for 7 of the satellites --
for example, the distance of And~II from M31 is increased by $81\%$ to
$270$ kpc. Positioning more of the satellites at larger radii requires
a larger, more massive halo to bind them.  Recently, Feast (1999) has
claimed that even the distance to M31 may be in serious error. He
suggested that the distance modulus for M31 is $24.75$ based upon a
critical comparison of several distance estimates obtained using a
variety of standard candles. This distance modulus implies a distance
to M31 of $\sim 900$ kpc. If this is the case, our mass estimate using
the standard priors rises by $28\%$ to $15.7 \times
10^{11}\Msun$. This is caused by the increased separation of the
satellites from M31 enforced by the greater distance. It is the most
substantial of all the changes recorded in Table~\resultsb.  Given the
size of recent revisions of the distances to the satellites, it seems
likely that the error bars quoted in Table~\satdata\ are rather
optimistic. We therefore re-analyse the data assuming error bars of
$25\%$ on all satellite distances in our error convolution. This has a
negligible effect on the mass estimate.

Next, we examine the effect of changing our assumed models for the
globular cluster and satellite number density laws. As discussed in
Section~3.1, choosing $\as = 250$ kpc provides a good fit to the
present satellite dataset, and leaves some room for more satellites to
be hiding within $250$ kpc of M31. Since we do not really know how
incomplete the satellite dataset is, our choice of $\as$ may be
incorrect. We therefore re-analyse the dataset with $\as = 150$
kpc. Table~\resultsb\ shows that the mass estimate changes by only
$6\%$. Similarly, choosing a larger value of $n$ for the globular
cluster number density law gives a better fit to the data at large
radii, while overestimating the numbers at small radii. If we choose
$n=5$, we find that the mass estimate changes by $\sim 12\%$. It is
thus clear that our choice of parameters for the number densities of
the tracer populations has little effect on the estimate obtained.
There is also some uncertainty as to the rotation velocity of the
outer globular cluster system (Elson \& Walterbos 1988; Huchra, Kent \& Brodie 1991), but we have verified that our results are robust to
changes in this parameter.

We must also consider the possibility that our assumption regarding
the normalisation of the halo potential is incorrect. Up to now, we
have assumed that the halo normalisation $\vc \sim 240$ \kms\ . If we
instead take $\vc \sim 270$ \kms, the mass estimate is reduced by
$\sim 9\%$, while if we reduce $\vc$ to $200$ \kms\ the mass is
slightly increased by $\sim 2\%$. The small size of the changes
confirms that this is not a real source of uncertainty.

\begintable*{\tabnumber}
\tabnam{\masses}
\def\tableunderline{\noalign{\vskip0.1truecm}\noalign{\vskip0.1truecm}}
\caption{{\bf Table \masses.}
Other mass estimates for M31 (out to a radius $r_{\rm max}$). All
estimates have been corrected to put M31 at a distance of 770 kpc by
applying the simple correction $\Delta M/M \sim \Delta D/D$ where
$\Delta M$ and $\Delta D$ are the changes in the mass estimate $M$ and
the distance $D$ to M31, respectively.}
\halign{#\hfil&\quad\hfil#\hfil&\quad\hfil#\hfil&\quad\hfil#\hfil&\quad\hfil#\hfil\cr
\noalign{\hrule}
\noalign{\vskip0.1truecm}
Author & Data & Mass ($10^{11} \Msun$) & $r_{\rm max}$ (kpc) \cr
\tableunderline
Rubin \& Ford (1970) & H$\alpha$ Rotation Curve & 2.0 & 27 \cr
\tableunderline
Einasto \& R\"ummel (1970) & Optical \& Radio data & 2.2 & $25$  \cr
\tableunderline
Gottesman \& Davies (1970) &  21cm Rotation Curve & 2.5 & 34\cr
\tableunderline
Hartwick \& Sargent (1970) & Globular Clusters & 3.8 & $ 19$ \cr
\tableunderline
Deharveng \& Pellet (1975) & H$\alpha$ Rotation Curve & 1.8 & 22 \cr
\tableunderline
Roberts \& Whitehurst (1975) & 21cm Rotation Curve & 4.1 & 33  \cr
\tableunderline
Gunn (1975) & Local Group Timing & 22 & -- \cr
\tableunderline
Hodge (1975) & Tidal cut-off of companions & 67 & Total  \cr
\tableunderline
Rood (1979) & Virial Mass of M31 subgroup & 2.9 & To IC 1613 \cr
\tableunderline
van den Bergh (1981) & Satellites & 8.7 & To LGS3 ($\sim 260$) \cr
 & Globular Clusters & 2.8 & 22 \cr
\tableunderline
Braun (1991) & 21cm Rotation Curve & 2.2 & 31 \cr
\tableunderline
Courteau \& van den Bergh (1999) & Satellites & 13.5 & To LGS3 ($\sim
260$)\cr
\tableunderline
This Paper &  Planetary Nebulae & 2.8 & $\sim 31$ \cr
           & Globular Clusters & 4.7 & $\sim 40$ \cr
           & All Satellites  & 12.3 & Total \cr
\tableunderline
\noalign{\vskip0.1truecm}\noalign{\hrule}
}
\tabnew
\endtable
\vskip 0.5truecm

Table~\masses\ summarises previous studies of the mass of M31 to allow
comparison with our results (c.f. Hodge 1992, chap. 8). All the mass
estimates have been corrected to a standard value of distance to M31
of 770 kpc.  As in the case of the Milky Way (see Fig.~13 of WE99),
the inferred mass of M31 has tended to increase during the past 30
years. Our estimate actually agrees well, both with the most recent
estimate of the total mass (Courteau \& van den Bergh 1999) and
estimates of the mass of the inner $20-30$ kpc (e.g. van den Bergh
1981). The agreement is partly fortuitous, as van den Bergh omits some
satellite galaxies we have included. It is interesting to compare the
results obtained using our algorithm with those which would be
obtained using, for example, the projected mass estimator of Bahcall
\& Tremaine (1981). Using this estimator with the recommended
multiplicative factor for an isotropic velocity distribution yields a
mass estimate of $11.5\times 10^{11}\Msun$, which agrees well with our
results. If we instead use the factor which is recommended for use in
the absence of information on the velocity distributions, we obtain
the somewhat larger mass of $17.2 \times 10^{11}\Msun$. 

\section{Error Analysis}
\beginfigure{\fignumber}
\fignam{\figerrhist}
\centerline{\psfig{figure=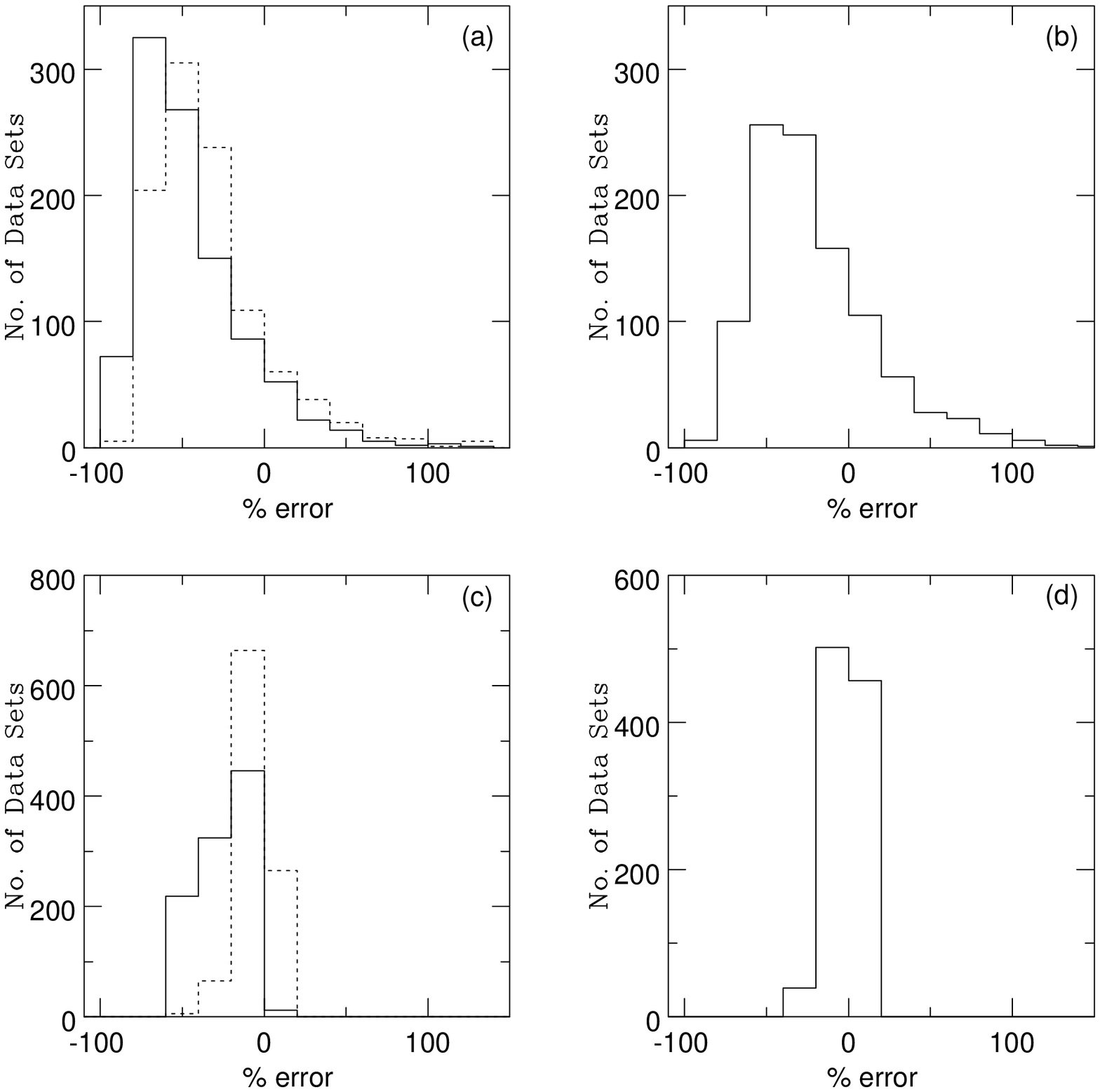,width=\hssize}}
\smallskip\noindent
\caption{{\bf Figure \figerrhist} (a) Histogram showing the spread in mass
estimates based on satellite datasets containing 10 objects (solid
line) and 20 objects (broken line) at radii $20$ kpc $< r < 500$ kpc
from M31. Each histogram shows the number of datasets out of 1000
which yielded a given percentage error in the total mass $M$.  (b)
Histogram illustrating the effects of incompleteness in the satellite
sample. Datasets of 20 satellites with different values of $\as$ used
in the generation of the data and the analysis. See text for
discussion. (c) Histogram showing the spread in mass estimates based
on globular cluster datasets containing 30 objects (solid line) and
100 objects (dotted line) at projected radii $20$ kpc $< R < 50$
kpc. The histogram shows the number of datasets out of 1000 which
yielded a given percentage error in the mass within $50$ kpc. (d)
Histogram illustrating the reduced bias in mass estimates based on
globular cluster datasets containing 100 objects at projected radii
$20$ kpc $< R < 50$ kpc for which proper motions have been obtained
using GAIA.}
\fignew
\endfigure
\noindent
This section examines the main sources of error in our estimate.  The
dominant problems are caused by the small size and possible
incompleteness of the dataset. The errors in the radial velocities are
mostly insignificant, the distances to the satellite galaxies being
the only quantities in the problem with large uncertainties.

\subsection{The Size of the Dataset}
\noindent
In order to investigate this problem, we generate 1000 synthetic
datasets of 10 satellites each, drawn from a TF model density
distribution with a DF of the form given in eq. (\anisoDF). The
anisotropy scale $\ra$ is taken as 166 kpc (which is the mean of the
prior probability distribution over the allowed range of 2.5 and 1000
kpc).  In carrying out this procedure, we use a Gaussian approximation
to the actual DF taking the velocity dispersions as the widths of the
Gaussians. This approximation can very occasionally lead to the
generation of satellites which are not bound -- these are rejected
from our datasets. Each dataset is analysed using the algorithm
described in Section~4 and the most likely value of the total mass of
the halo is noted.

The solid histogram in Fig.~\figerrhist(a) shows the number of
datasets which yield a given percentage error in the mass
estimate. There is a tendency to underestimate the mass with $\sim
50\%$ of estimates more than a factor of two smaller than the true
value. Only $\sim 14\%$ of estimates are within $20\%$ of the correct
value. There is also a large spread in mass estimates -- the average
absolute deviation of the estimates about the mean is $49\%$.  It is
reasonable to hope that the number of satellite galaxies could rise to
perhaps 20 in the near future, given the successes of the ongoing
surveys of the sky around Andromeda (Armandroff \& Da Costa 1999;
Karachentsev \& Karachentseva 1999).  The dotted histogram of
Fig.~\figerrhist(a) shows the effect of this increase -- the
systematic underestimate is still present although it is somewhat less
than that for the 10 satellite case.  We conclude that there is a
systematic tendency to underestimate the mass by a factor of two due
solely to the size of the dataset. This is coupled with a spread of
order $\sim 50\%$.

As the number of datapoints becomes large, our prior probabilities
become unimportant. With small datasets, our choice for priors can
lead to systematic biases. WE99 suggested that this effect was
significant for the Milky Way, as the customary priors tend to be
biased towards radial anisotropy, while the dataset, at least judging
from the available proper motions, is not. For the case of the
simulations reported in Fig.~\figerrhist(a), the mean of the mass
prior $1/M^2$ is less than the true mass, which partly accounts for
the effect. However, this circumstance is only slightly improved if we
switch to a mass prior of $1/M$, whose mean does correspond to the
true mass.  There is clearly scope for more work on the possible
causes of bias of mass estimators with small datasets. Both
underestimates and overestimates have been reported by others using
virial mass estimators (e.g., Haller \& Melia 1996, Aceves \& Perea
1999). It seems likely that Bayesian estimators can also yield biased
results of either sign, depending on the details of the potential,
distribution function and priors. However, for our models, small
datasets of radial velocities typically lead to underestimates.
\subsection{Incompleteness}
\noindent
In order to study the effects of incompleteness, we generate datasets
of 20 satellites in which the positions are chosen according to a TF
profile with $\as = 150$ kpc, while the velocities are chosen assuming
a value of $\as = 300$ kpc in the number density profile. When
analysing the data, a value of $\as = 150$ kpc is assumed. This is
analogous to the observational situation in which the observed number
density distribution may not correspond to the true distribution if
the sample is incomplete. The measured velocities are however governed
by the true velocity distribution.

Fig.~\figerrhist(b) presents the results of performing 1000 such
simulations. Comparing this histogram with the dotted histogram of
Fig.~\figerrhist(a), we see that the incompleteness has not degraded
our ability to estimate the mass in any significant way. We therefore
conclude that the (unknown) incompleteness of our satellite sample
should not bias our mass estimate, provided that the incompleteness is
unrelated to the kinematics of the satellites. However, one can
imagine cases in which this could happen -- for example, if there is a
systematic tendency for the fainter satellites to move on different
types of orbits than the brighter satellites. In this case, the
magnitude limit enforces a kinematic bias which is difficult to model,
but could have a deleterious effect on the resulting mass
estimates. However, if the only incompleteness is due to magnitude
limits in our surveys and is kinematically unbiased, then
Fig.~\figerrhist(b) shows us that the major worry is the small number
of satellites and not the incompleteness of the sample.
\subsection{Globular Cluster Datasets}
We now examine the uncertainties in our mass estimates for the inner
region of the M31 halo based on the globular clusters and planetary
nebulae. Bearing in mind the ongoing searches (Perrett et al. 1999),
our main motivation is to discover how many radial velocities are
needed for reliable results.  The solid histogram of
Fig.~\figerrhist(c) shows the results of estimates of the mass inside
$50$ kpc based on simulated datasets of 30 globular clusters with
projected radii $20$ kpc $\lta R \lta 50$ kpc drawn from a rotating DF
of the form (\frot). As in the case of the satellites, there is a
systematic bias towards underestimates of the mass, although this is
less pronounced than the factor of two bias observed in the satellite
case. The spread in mass estimates is also reassuringly small ($\sim
15\%$). The dotted histogram in Fig.~\figerrhist(c) shows the results
of simulations of datasets consisting of 100 globular clusters at
projected radii out to $50$ kpc. This histogram demonstrates the value
of searching for globular clusters orbiting M31 at larger radii. The
systematic bias in the mass estimate is greatly reduced and $\sim
90\%$ of the estimates lie within $10\%$ of the true value. The mass
within $30$ kpc can also be recovered to excellent accuracy ($\sim
93\% $ of estimates lie within $10\%$ of the true answer).
\beginfigure{\fignumber}
\fignam{\figsimhist}
\centerline{\psfig{figure=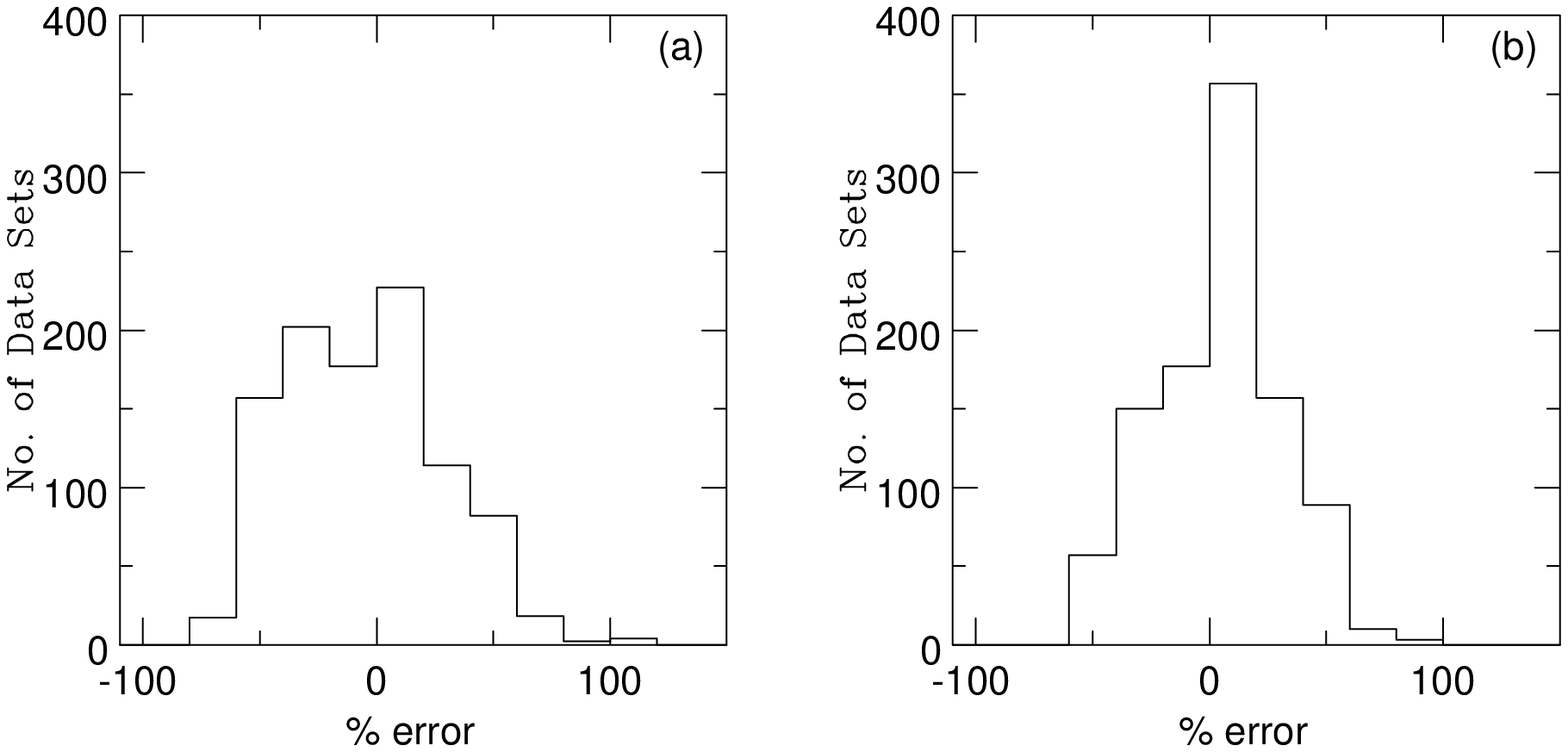,width=\hssize}}
\smallskip\noindent
\caption{{\bf Figure \figsimhist} (a) Histogram showing the spread in mass
estimates based on satellite datasets containing 10 objects at radii
$20$ kpc $< r < 500$ kpc whose proper motions have been measured using
SIM and GAIA. The histogram shows the number of datasets out of 1000
which yielded a given percentage error in the total mass $M$. (b) As
in (a), but for datasets of 20 satellites.}
\fignew
\endfigure
\section{Conclusions and Prospects}
Our best estimate for the mass of the M31 halo based on the motions of
satellite galaxies, planetary nebulae and globular clusters is $12.3
\times 10^{11}\Msun$. There is at least a factor of two
uncertainty in this value due to the small sizes of the three tracer
datasets. Using Monte Carlo simulations of datasets of 10 satellites
we have shown that there is a tendency to underestimate the mass with
this algorithm when the data are very sparse -- we find that $50\%$ of
mass estimates underestimate the mass by more than a factor of
two. There is also a spread in mass estimates of order
$50\%$. Incorporating these into error bars on our mass estimate, we
obtain $12.3^{+18}_{-6} \times 10^{11}\Msun$.  For comparison, our
earlier estimate for the Milky Way halo (WE99) was $19^{+36}_{-17}
\times 10^{11}\Msun$. Though the error bars are large, we reach the
surprising conclusion that {\it the total mass of M31 may well be less
than that of the Milky Way}.

This is contrary to current opinion. Almost all recent authors have
argued that M31 is the most massive member of the Local Group (e.g.,
Peebles 1989; Hodge 1992; Mateo 1998; Courteau \& van den Bergh 1999).
These authors adduce a number of pieces of evidence to support their
viewpoint. The asymptotic value of the rotation curve of M31 seems to
be $\sim 10 \%$ higher than that of the Milky Way (Rubin \& Ford
1970). The number of globular clusters in M31 is more than double that
in the Milky Way (Fusi Pecci et al. 1993). The scalelength of the M31
disk exceeds that of the Milky Way (Walterbos \& Kennicutt 1988). The
face-on B-band absolute magnitude of M31 is $M_{\rm B} = -21.1 \pm
0.4$, whereas that of the Milky Way is $M_{\rm B} = -20.5 \pm 0.5$
(Hodge 1992).  On the other hand, the infrared luminosity of the Milky
Way is much higher than that of M31 (Walterbos \& Schwering 1987). The
mass in hydrogen gas in the Milky Way also exceeds that of M31 (Hodge
1992).

The difficulty in all this is judging whether any of these statements
are properly comparing like with like. M31 has Hubble type Sb, whereas
the Hubble type of the Milky Way is certainly later, perhaps Sbc or Sc
(e.g., Gilmore, King \& van der Kruit 1989). On such grounds alone, it
is natural to expect that M31 has a larger bulge than the Milky
Way. Late-type spirals are gas-rich and so it is also natural to
expect the mass in gas in the Milky Way to exceed that of M31. The
relative number of globular clusters is a poor argument, as it is only
weakly correlated with total mass (e.g., Ashman \& Zepf 1998).  There
is also no correlation between the satellite velocity dispersion in
the outer halo of a galaxy and the rotation velocity of its disk
(Zaritsky, Smith, Frenk \& White 1997). Thus, a higher disk rotation
velocity does not necessarily imply a more massive halo.

We believe that the mass within $\sim 30$ kpc is greater for Andromeda
than for the Milky Way, though probably only by about $20\%$. The best
guide to total mass is provided by the kinematics of distant tracers
like the satellite galaxies. Our analysis of both galaxies is the most
complete and sophisticated to date, and it suggests that the total
mass of the M31 halo is probably less than that of the Milky
Way. Beyond $\sim 30$ kpc, the mass to light ratio of M31 seems to
increase more slowly than in the Milky Way.

There is a pressing need for more data on objects in the halo of
M31, if we are to consolidate this result. There are a number of lines
of attack.  First, there are five satellites of M31 whose radial
velocities have not yet been measured, namely And~I, And~III, And~V,
Pegasus~II (And~VI) and Cassiopeia (And~VII). The measurement of the
radial velocities of these galaxies is being carried out at present
(Grebel 2000, private communication). This will increase by 50 percent
the number of tracer objects at large radii and it will be interesting
to see how the mass estimate changes. There is even cause for optimism
that the dataset will increase in size. The current rate of discovery
of dwarf spheroidals of M31 (And~V, And~VI and And~VII were discovered
in 1998) emphasises the number of satellites which have thus far been
missed. Karachentsev \& Karachentseva (1999) state that the
North-South asymmetry in the distribution of satellites about M31
could imply that 3-5 satellites of M31 have escaped detection due to
galactic extinction and the extended cirrus fields in the vicinity of
M31. A second line of attack is the discovery and exploitation of halo
globular clusters and planetary nebulae (Bridges 1999; Perrett et
al. 1999).  Based on a comparison with the Milky Way globular cluster
system, Fusi Pecci et al. (1993) estimate that outside $20$ kpc, there
are likely to be $\sim 30$ globular clusters brighter than
V=18. Identifying these distant bright clusters and measuring their
radial velocities is an important target for constraining the total
mass. The surveys in progress are concentrated in the central regions
and contain only a handful of objects outside $\sim 20$ kpc (Perrett
1999, private communication).  These are primarily useful for
measuring the mass distribution in the inner parts -- for example, a
dataset of 100 globular clusters with radial velocities would allow us
to determine the mass within $30$ kpc to within $10\%$ (see
Fig.~\figerrhist(c)). The number of planetary nebulae in M31 is known
to be of the order of $10^4$ (Nolthenius \& Ford 1987). Many hundreds
of planetary nebulae lie in the halo and are invaluable as probes of
the potential in regions where no other tracers have been found.
Nowadays, these can be identified relatively easily with wide-field
imaging using OIII and continuum filters and their velocities can be
found with multi-object spectroscopy. So, there are real prospects
that much larger datasets will be available soon.

Within the decade, astrometric satellites will allow the determination
of the proper motions of some of the globular clusters of M31. While
this is beyond the reach of current ground based instruments, the
proposed {\it Global Astrometric Interferometer for Astrophysics}
(GAIA) satellite will scan the whole sky and will perform astrometry
on all objects brighter than $V=20$. The apparent magnitudes of the
unresolved cores of M31's bright globular clusters are typically
around $V \approx 16-17$.  Fig.~\figerrhist(d) shows how the mass
estimate within $50$ kpc is affected by the addition of proper motion
measurements for a dataset of $100$ globular clusters with $20<R<50$
kpc. The systematic bias visible in the solid histogram of
Fig.~\figerrhist(c) when only radial velocities are used is entirely
absent when proper motion data are available and $96\%$ of mass
estimates lie within $\pm 20\%$ of the true mass. The mass inside $30$
kpc is recovered with even greater confidence, as $96\%$ of estimates
lie within $10\%$ of the true mass.

It will also be possible to obtain proper motions for some or all of
the M31 dwarf galaaxy satellites. In this case, the proposed {\it
Space Interferometry Mission} (SIM) satellite is perhaps the most
appropriate instrument to use. SIM is a pointing instrument. It will
study fewer objects than GAIA, but with greater accuracy.  SIM is
planned to have an astrometric precision of $2 \mu$as yr$^{-1}$ for
objects with V=20. Almost all the Local Group members have at least
some stars with magnitudes brighter than $V=20$, making proper motion
determinations feasible. The brighter objects, like M32 and M33, have
$\sim 10^4$ stars with $V < 20$, while even the fainter ones, like IC
10 and NGC 147, have $\sim 10$.  At the distance of M31, a proper
motion uncertainty of $2 \mu$as yr$^{-1}$ corresponds to an error in
the velocity transverse to the line of sight of $\sim 7$\kms.
Fig.~\figsimhist(a) shows how the systematic underestimate is removed
even if we only measure the proper motions of the 10 satellites which
currently have radial velocities. The spread in mass estimates is
still large (the average absolute deviation about the mean is $\sim
27\%$ of the true mass estimate), as there are still only 10
datapoints. In the optimistic case in which the satellite dataset
swells to 20 and a combination of SIM and GAIA obtain all the proper
motions, Fig.~\figsimhist(b) shows that the mass estimates are much
more strongly peaked around the true answer and the spread is reduced
to $\sim 22\%$ of the true mass.

The masses of the Milky Way and the Andromeda galaxies are fundamental
parameters that we need to know to understand the Local Group.
Although our investigations have convinced us that the masses are
rather imperfectly known at present, there are excellent prospects for
improvement in the very near future.

\section*{Acknowledgments}
NWE thanks the Royal Society for financial support, while MIW
acknowledges help from a Scatcherd Scholarship. We thank Tim de Zeeuw
and Gerry Gilmore for information on the GAIA satellite, as well as
James Binney and Scott Tremaine for useful advice. We are grateful to
Terry Bridges, Dave Carter and Kathy Perrett for keeping us informed
of their globular cluster and planetary nebulae surveys of M31. Kathy
Perrett kindly fowarded data on the positions and radial velocities of
the outermost globular clusters in advance of publication.

\appendix
\section{A Regularising Transformation}

As noted earlier, DFs of the form $l^{-2\beta}f(\varepsilon)$ present
a difficulty when calculating the probabilities (\satprob) or
(\projprob) since, for $\beta > 0$, they possess a singularity at
$l=0$. This singularity, however, is integrable for all $\beta <1$.
Van der Marel et al. (2000) refer to this problem and use the
substitution $s = \ln \tan \eta$ to remove the singularity.  (We
recall from Section 4.1 that ($\vT, \eta$) are polar coordinates in
the plane of projection).  This substitution solves the problem for $0
< \beta < 1$, but only in the case of the projected probability
distributions (\projprob).

For the unprojected probability distributions (\satprob), van der
Marel et al.'s (2000) substitution regularises the integral for $\beta
< 1/2$.  In order to calculate the probability (\satprob) numerically
for $1/2 < \beta <1$, we have to make the additional transformation in
velocity space $(\vT, \eta) \rightarrow (\lambda, \mu)$, where
$$\eqalign{\lambda &= [\vT \cos (\theta_i + \alpha_i) + 
\vlosic \sin (\theta_i + \alpha_i)]^{2-2\beta}\cr
\mu &= \arctan \left( {\eta \vT \over \vT \cos (\theta_i + \alpha_i) +
\vlosic \sin (\theta_i + \alpha_i)}\right)}\eqno\new$$
in the region of the singularity.


\section*{References}
 
\beginrefs

\bibitem Aceves H., Perea J., 1999, A\&A, 345, 439

\bibitem Armandroff T. E., Da Costa G. S., 1999, in ``The Stellar
Content of Local Group Galaxies'', Whitelock, P. \& Cannon, R. eds.,
IAU Symp. Series, 192, 203

\bibitem Arnaboldi M., Napolitano N., Capaccioli M., 2000, 
in ``Galaxy Dynamics: From the Early Universe to the Present'',
eds. Combes F., Mamon G., Charmandaris V., ASP. Conf. Ser., 197, 103

\bibitem Ashman K., Zepf S., 1998, ``Globular Cluster Systems'',
Cambridge University Press, Cambridge

\bibitem Bahcall J. N., Tremaine S. D., 1981, ApJ, 244, 805

\bibitem Barmby P., Huchra J. P., Brodie J. P., Forbes D. A., Schroder L. L.,
Grillmair C. J., 2000, AJ, 119, 727

\bibitem Binney J., Tremaine S., 1987, ``Galactic Dynamics'', Princeton
University Press, Princeton

\bibitem Braun R., 1991, ApJ, 372, 54

\bibitem Bridges T., 1999, in ``Galaxy Dynamics - A
Rutgers Symposium'', Merritt D., Sellwood J. A., Valluri M. eds., ASP
Conf. Ser., 182, 415

\bibitem C$\hat{\rm o}$t$\acute{\rm e}$ P., Mateo M., Olszewski
E. W., Cook K. H., 1999, ApJ, 526, 147

\bibitem Courteau S., van den Bergh S., 1999, AJ, 118, 337

\bibitem Crampton D., Cowley A., Schade D., Chayer P., 1985, ApJ, 288,
494

\bibitem Crotts A. P. S., 1992, ApJ, 399, L43

\bibitem Crotts A. P. S., Tomaney A., 1996, ApJ, 473, L87

\bibitem Deharveng J. M., Pellet A., 1975, A\&A, 38, 15

\bibitem Djorgovski S., Meylan G., 1994, AJ, 108, 1292

\bibitem Einasto J., R\"ummul U., 1970, in ``The Spiral Structure of
our Galaxy'', Becker W., Contopoulos G., eds., IAU
Symp. 38, 51

\bibitem Elson R. A. W., Walterbos R. A. M., 1988, ApJ, 333, 594

\bibitem Evans N. W., 1993, MNRAS, 260, 191

\bibitem Evans N. W., H\"afner R. M., de Zeeuw P.T., 1997, MNRAS, 286,
315

\bibitem Evans N. W., Wilkinson M. I., 2000, in ``Microlensing 2000: A
New Era of Microlensing Astrophysics'', Menzies J. W., \& Sackett
P. D., eds, ASP. Conf. Ser., in press

\bibitem Feast M., 1999, in "The Stellar Content of Local Group
Galaxies", Whitelock P. \& Cannon R. eds., IAU Symp. Ser., 192, 203

\bibitem Federici L., B\`onoli F., Ciotti L., Fusi Pecci F., Marano
B., Lipovetsky V.A., Neizvestny S.I., Spassova N., 1993, A\&A, 274,
87

\bibitem Ford H. C., Ciardullo R., Jacoby G. H., Hui X., 1989,
in ``Planetary Nebulae'', ed. Torres-Peimbert S., IAU Symp. Series, 131,
335 (Kluwer, Dordrecht)

\bibitem Fusi Pecci F., Cacciari L., Federici L., Pasquali A.,
1993, in ``The Globular Cluster-Galaxy Conncection'', Smith G. H. \&
Brodie J. P. eds., ASP. Conf. Ser. 48, 410

\bibitem Gilmore G., King I. R., van der Kruit P.C., 1989, ``The Milky
Way as a Galaxy'', University Science Books, Mill Valley, California

\bibitem Gottesman S. T., Davies R. D., 1970, MNRAS, 149, 263

\bibitem Grebel E. K., 1999, in ``The Stellar Content of Local
Group Galaxies'', Whitelock, P. \& Cannon, R. eds., IAU
Symp. Ser. 192, 17

\bibitem Gunn J., 1975, Comments Ap. Sp. Phys., 6, 7

\bibitem Hartwick F., Sargent W., 1974, ApJ, 190, 283

\bibitem Haller J.W., Melia F., 1996, ApJ, 464, 774 

\bibitem Heisler J., Tremaine S. D., Bahcall J. N., 1985, ApJ, 298, 8

\bibitem H\'enon M., 1973, A\&A, 24, 229

\bibitem Hodge P., 1975, BAAS, 7, 506

\bibitem Hodge P., 1992, ``The Andromeda Galaxy'', Kluwer Academic,
Dordrecht

\bibitem Huchra J. P., Kent S. M., Brodie J. P., 1991, ApJ, 370, 495

\bibitem Huchra J. P., Stauffer J., van Speybroeck L., 1982, ApJ, 259, L57

\bibitem Kahn F., Woltjer L., 1959, ApJ, 130, 705

\bibitem Karachentsev I. D., Makarov D. I., 1996, AJ, 111, 794

\bibitem Karachentsev I. D., Tikhonov N. A., Sazonova L. N., 1994,
Astr. Letters, 20, 84

\bibitem Karachentsev I. D., Karachentseva V. E., 1999, A\&A, 341, 355

\bibitem Kendall M., Stuart A., 1977, The Advanced Theory of
Statistics, Griffin, London

\bibitem Kent S. M., Huchra J. P., Stauffer J., 1989, AJ, 98, 2080

\bibitem Kerins E. J., Carr B. J., Evans N. W., Hewett P., Lastennet
E., Le Du Y., Melchior A.-L., Smartt S., Valls-Gabaud D., 2000, MNRAS,
submitted (astro-ph/0002256)

\bibitem Kochanek C., 1996, ApJ, 457, 228

\bibitem Kulessa A. S., Lynden-Bell D., 1992, MNRAS, 255, 105

\bibitem Lee M. G., Freedman W. L., Madore B. F. 1993, in ``New 
Perspectives on Stellar Pulsation and Pulsating Variable Stars'',
eds. J. M. Nemec, J. M. Matthews (Cambridge: Cambridge Univ. Press),
92

\bibitem Little B., Tremaine S. D., 1987, ApJ, 320, 493

\bibitem Lynden-Bell D., 1999, in ``The Stellar Content of Local
Group Galaxies'', Whitelock, P. \& Cannon, R. eds., IAU
Symp. Ser. 192, 39

\bibitem Mateo M., 1998, ARA\&A, 36, 435

\bibitem Merritt D., 1985, AJ, 90, 1027

\bibitem Nolthenius R., Ford H. C., 1987, ApJ, 317, 62

\bibitem Osipkov L. P., 1979, Pis'ma Astr. Zh., 5, 77

\bibitem Peebles P. J. E., 1989, ApJ, 344, L53

\bibitem Perrett K., Hanes D., Bridges T., Irwin M., Carter D.,
Huchra J., Brodie J., Watson F., 1999, in ``Galaxy Dynamics - A
Rutgers Symposium'', Merritt D., Sellwood J. A., Valluri M. eds.,
ASP Conf. Ser., 182, 431


\bibitem Roberts M., Whitehurst R., 1975, ApJ, 201, 327

\bibitem Rood H. J., 1979, ApJ, 232, 699

\bibitem Rubin V., Ford W., 1970, ApJ, 159, 379

\bibitem Sargent W. L. W., Kowal C. T., Hartwick F. D. A., van den
Bergh S., 1977, AJ, 82, 947

\bibitem Schmoldt I., Saha P., 1998, AJ, 115, 2231

\bibitem van den Bergh S., 1981, PASP, 93, 428

\bibitem van den Bergh S., 1998, AJ, 116, 1688

\bibitem van den Bergh S., 1999a, AJ, 117, 2211

\bibitem van den Bergh S., 1999b, A\&AR, 9, 273

\bibitem van der Marel R. P., Magorrian J., Carlberg R. G., Yee
H. K. C., Ellingson E., 2000, AJ, submitted, astro-ph/9910494 

\bibitem Walterbos R. A. M., Kennicutt R. C., 1988, A\&A, 198, 61

\bibitem Walterbos R. A. M., Schwering P. B. W., 1987, A\&A, 180, 27

\bibitem Wilkinson M. I., 1999, Ph.D. thesis, Oxford University

\bibitem Wilkinson M. I., Evans N. W., 1999, MNRAS, 310, 645 (WE99) 

\bibitem Zaritsky D., Smith R., Frenk C. S., White S. D. M., 1997,
ApJ, 478, 39

\eqnumber =1
\def\chaphead{\hbox{A}}
\endrefs
\bye